\documentclass[11pt]{article}

\usepackage[margin=0.88in,headheight=22pt]{geometry}
\usepackage[T1]{fontenc}
\usepackage{lmodern}
\usepackage{microtype}
\usepackage{graphicx}
\usepackage[table,dvipsnames]{xcolor}
\usepackage{xurl}
\usepackage{array}
\usepackage{longtable}
\usepackage{booktabs}
\usepackage{tabularx}
\usepackage{ragged2e}
\usepackage{amsmath}
\usepackage{amssymb}
\usepackage{caption}
\usepackage{enumitem}
\usepackage{etoolbox}
\usepackage{float}
\usepackage{fancyhdr}
\usepackage{titlesec}
\usepackage{needspace}
\usepackage{listings}
\usepackage[most]{tcolorbox}
\usepackage[mathlines]{lineno}

\definecolor{ReportInk}{HTML}{1F2933}
\definecolor{ReportMuted}{HTML}{5B6572}
\definecolor{ReportBlue}{HTML}{2563A6}
\definecolor{ReportTeal}{HTML}{138A8A}
\definecolor{ReportOrange}{HTML}{B86B17}
\definecolor{ReportPanel}{HTML}{F6F8FA}
\definecolor{ReportSoft}{HTML}{EEF7F6}
\definecolor{ReportLine}{HTML}{D7DEE8}
\definecolor{ReportCodeBg}{HTML}{F7F4EF}
\definecolor{ReportCodeFrame}{HTML}{E8D9C2}
\usepackage[
  colorlinks=true,
  linkcolor=ReportBlue,
  citecolor=ReportTeal,
  urlcolor=ReportBlue
]{hyperref}

\newcommand{\system}{\textsc{CLI-Anything}}
\newcommand{\hub}{\textsc{CLI-Hub}}
\newcommand{\jsonflag}{\texttt{--json}}
\newcommand{\skill}{\texttt{SKILL.md}}

\renewcommand{\headrulewidth}{0.4pt}
\renewcommand{\headrule}{\hbox to\headwidth{\color{ReportLine}\leaders\hrule height \headrulewidth\hfill}}

\titleformat{\section}[block]
  {\Large\sffamily\bfseries\color{ReportInk}}
  {\color{ReportTeal}\thesection}{0.75em}{}
  [\vspace{0.28em}{\color{ReportTeal!35}\titlerule[0.7pt]}]
\titleformat{\subsection}
  {\large\sffamily\bfseries\color{ReportInk}}
  {\color{ReportOrange}\thesubsection}{0.7em}{}
\titleformat{\subsubsection}
  {\normalsize\sffamily\bfseries\color{ReportInk}}
  {\color{ReportTeal}\thesubsubsection}{0.65em}{}
\titlespacing*{\section}{0pt}{1.75em plus 0.25em minus 0.15em}{1.0em}
\titlespacing*{\subsection}{0pt}{1.15em plus 0.2em minus 0.1em}{0.55em}
\titlespacing*{\subsubsection}{0pt}{0.95em plus 0.15em minus 0.1em}{0.42em}
\pretocmd{\section}{\Needspace{8\baselineskip}}{}{}
\pretocmd{\subsection}{\Needspace{5\baselineskip}}{}{}
\pretocmd{\subsubsection}{\Needspace{4\baselineskip}}{}{}

\setlist[enumerate]{
  leftmargin=*,
  itemsep=0.18em,
  topsep=0.2em,
  label=\textcolor{ReportTeal}{\arabic*.}
}

\newcolumntype{Y}{>{\RaggedRight\arraybackslash}X}
\newcolumntype{Z}{>{\RaggedLeft\arraybackslash}X}
\newcommand{\tablehead}{\rowcolor{ReportTeal!10}}

\lstdefinestyle{reportcode}{
  basicstyle=\ttfamily\small\color{ReportInk},
  breaklines=true,
  columns=fullflexible,
  keepspaces=true,
  showstringspaces=false,
  keywordstyle=\color{ReportBlue}\bfseries,
  commentstyle=\color{ReportMuted}\itshape,
  stringstyle=\color{ReportOrange}
}

\newtcblisting{codeblock}[1][]{
  enhanced,
  breakable,
  listing only,
  listing options={style=reportcode},
  title={#1},
  colback=ReportCodeBg,
  colframe=ReportCodeFrame,
  coltitle=ReportInk,
  fonttitle=\sffamily\bfseries\small,
  boxrule=0.45pt,
  arc=1.5mm,
  left=2.5mm,
  right=2.5mm,
  top=1.5mm,
  bottom=1.5mm,
  borderline west={2pt}{0pt}{ReportOrange}
}

\newcommand{\framedgraphic}[2][0.92\linewidth]{%
  \begin{tcolorbox}[
    enhanced,
    colback=white,
    colframe=ReportLine,
    boxrule=0.45pt,
    arc=1.5mm,
    left=1.2mm,
    right=1.2mm,
    top=1.2mm,
    bottom=1.2mm,
    drop fuzzy shadow=ReportLine!55
  ]
  \centering\includegraphics[width=#1]{#2}
  \end{tcolorbox}%
}

\newcommand{\MetricHarnessEntries}{65}
\newcommand{\MetricPublicEntries}{18}
\newcommand{\MetricCombinedEntries}{83}
\newcommand{\MetricHarnessCategories}{29}
\newcommand{\MetricPublicCategories}{11}
\newcommand{\MetricCombinedCategories}{32}

\newcommand{\MetricSkillFiles}{61}

\newcommand{\MetricPreviewHelpers}{5}

\newcommand{\MetricAgentShare}{87.8\%}

\newcommand{\MetricAgentRatio}{7.19$\times$}

\newcommand{\MetricBlenderGroups}{12}
\newcommand{\MetricBlenderCommands}{54}
\newcommand{\MetricBlenderPublicCommands}{53}
\newcommand{\MetricBlenderHiddenCommands}{1}

\newcommand{\MetricBlenderModifiers}{8}

\newcommand{\MetricBlenderPreviewRecipes}{1}
\newcommand{\MetricStsTwoCliCommands}{28}
\newcommand{\MetricStsTwoDecisionStates}{15}
\newcommand{\MetricStsTwoBridgeActions}{24}

\newcommand{\MetricStsTwoUnitTests}{9}
\newcommand{\MetricStsTwoEndToEndTests}{5}

\title{\system: Towards Agent-Native Computer Use}

\author{
Yuhao Yang, Tianyu Fan, Chao Huang\\
University of Hong Kong\\
}

\makeatletter
\renewcommand{\maketitle}{%
  \thispagestyle{empty}
  \begin{tcolorbox}[
    enhanced,
    colback=ReportPanel,
    colframe=ReportLine,
    boxrule=0.45pt,
    arc=2mm,
    left=6mm,
    right=6mm,
    top=6mm,
    bottom=6mm,
    borderline west={4pt}{0pt}{ReportTeal},
    drop fuzzy shadow=ReportLine!65
  ]
    {\sffamily\bfseries\fontsize{19}{23}\selectfont\color{ReportInk}\@title\par}
    \vspace{0.75em}
    {\small\color{ReportMuted}\@author\par}
  \end{tcolorbox}
}
\makeatother

\renewenvironment{abstract}
  {\begin{tcolorbox}[
    enhanced,
    breakable,
    title={Abstract},
    colback=ReportSoft,
    colframe=ReportTeal!45,
    coltitle=ReportInk,
    fonttitle=\sffamily\bfseries,
    boxrule=0.55pt,
    arc=2mm,
    left=4mm,
    right=4mm,
    top=2mm,
    bottom=2mm
  ]\small}
  {\end{tcolorbox}}

\begin{document}

\maketitle

\begin{abstract}
As large language models advance in reasoning and tool use capabilities, researchers increasingly seek to leverage them for computer use agents that can interact with existing software. The dominant approach develops GUI agents that control applications through visual interfaces—interpreting screenshots, locating UI elements, and executing mouse clicks to mimic human interaction. This GUI-centric paradigm fundamentally misaligns with agent capabilities. Current GUI agents struggle with brittle pixel-level interactions, timing dependencies, and coordinate-based actions that break with interface changes. They force agents to emulate human perceptual limitations rather than leverage their computational strengths in structured data processing and programmatic control. CLI-Anything argues for agent-native computer use design. Instead of forcing agents to navigate visual layouts, we create interfaces aligned with how agents naturally operate—through structured commands, explicit state representations, and deterministic feedback. We transform existing applications into command-line harnesses that preserve functionality while exposing machine-readable protocols optimized for AI-native interaction. This eliminates the lossy visual-to-computational translation that plagues GUI agents. Rather than building sophisticated screen readers and click simulators, we should redesign interaction paradigms around agent strengths: precise programmatic control and deterministic execution. We examine the methodology, architecture, evidence, and future directions for this agent-native transformation of computer use. We have built CLI-Hub as a comprehensive platform that operationalizes this agent-native computer use vision. The platform provides methodology, architecture, and infrastructure for this fundamental transformation of computer use.\\

CLI-Anything is available at: \href{https://github.com/HKUDS/CLI-Anything}{https://github.com/HKUDS/CLI-Anything}.\\
CLI-Hub is available at: \href{https://clianything.cc}{https://clianything.cc}.
\end{abstract}


\section{Introduction}

Agents are increasingly evaluated in settings that require tool use, long-horizon
state, executable feedback, and artifact construction. Early agent work made
reasoning-action loops and tool calls explicit~\cite{react,toolformer}. API
systems and benchmarks made tool selection and argument construction
measurable~\cite{toolllm,gorilla,apibank}. Coding-agent environments then made
success observable through repository tests~\cite{swebench,sweagent}. Web and
desktop environments extended the same pressure to task-state
checks~\cite{webarena,osworld}. Work on reflection and open-ended agents shows why persistent memory
and recovery matter over long trajectories~\cite{reflexion,voyager}. Recent
agentic model work follows the same direction: Qwen3-Coder-Next frames coding
agents around verifiable tasks paired with executable
environments~\cite{qwen-coder-next}; DeepSeek-V3.2 highlights large-scale
agentic task synthesis and tool-use training~\cite{deepseek-v32}; and Kimi K2.5
introduces multimodal agentic workflows and a parallel Agent
Swarm~\cite{kimi-k25}. Together, these works point toward an operational
constraint for agent systems: agents learn and improve when work is grounded in
environments that can execute, check, and return structured feedback.

\system{} approaches the same constraint as a foundational software-interface
problem. It asks what interface real software should expose to agents once
clicking through the application becomes the wrong abstraction. The core claim is
simple:
software artifacts should be created, inspected, rendered, verified, and replayed
through the same explicit boundary whenever that boundary exists. A GUI may
remain useful as a fallback and as a human display surface. For many LLM-agent
workflows, it is the wrong default control surface.

GUIs are well designed for humans~\cite{directmanipulation}. They compress state into pixels, rely on
pointing and focus, hide the underlying project graph, and assume continuous
perception. Agents need different
properties: stable verbs, explicit state, JavaScript Object Notation (JSON)
output, durable history,
programmatic validation, installable tools, and a way to discover the right
interface at task time. A stateful CLI harness acts as an agent-facing contract
that delegates final truth to the real application whenever possible.

This report contributes four things. First, it states the interface thesis behind
\system{}: professional applications need agent-facing contracts that sit above
their real backends. Second, it grounds the thesis in implemented system
evidence: harness generation, preview protocol, skill generation, and \hub{}
distribution.
Third, it situates \system{} within related work on verifiable environments and
tool-use training~\cite{qwen-coder-next,deepseek-v32,apibank}. It also connects
to executable app benchmarks and multimodal agentic workflows~\cite{appworld,
taubench,kimi-k25}: those workloads require software interfaces that are
more structured than pixels. Fourth, it identifies
limits and future work: broader preview coverage, agent-first \hub{} APIs,
an experimental scenario-level CLI Matrix, and benchmarkable artifact tasks.

\begin{table}[t]
\centering
\footnotesize
\begin{tabularx}{\linewidth}{>{\RaggedRight\arraybackslash}p{0.24\linewidth}>{\RaggedRight\arraybackslash}p{0.31\linewidth}Y}
\toprule
\tablehead
\textbf{Paper term} & \textbf{Repository location} & \textbf{Role in the system} \\
\midrule
Harness-generation SOP & \path{cli-anything-plugin/HARNESS.md} & Method for lifting GUI applications into stateful CLIs. \\
Shared read-eval-print loop (REPL) shell & \path{cli-anything-plugin/repl_skin.py} & Common terminal interaction layer copied into generated harnesses. \\
Generated harness & \path{<software>/agent-harness/} & Independent Python package for one lifted application. \\
Blender harness & \path{blender/agent-harness/cli_anything/blender/} & Case-study implementation used for scene, render, preview, and artifact validation. \\
Backend execution wrapper & \path{.../utils/<software>_backend.py} & Locates the real application, invokes it, and checks backend-produced artifacts. \\
Scene-contract lowering pass & \path{blender/.../utils/bpy_gen.py} & Converts the Blender JSON scene contract into executable \texttt{bpy}. \\
Backend-gated end-to-end (E2E) suite & \path{.../tests/test_full_e2e.py} & Exercises real software, subprocess entry points, native files, and rendered outputs. \\
Companion skills for CLIs & \path{skills/} & Agent-readable usage surfaces generated after the harness and preview contract are known. \\
\hub{} installer & \path{cli-hub/cli_hub/installer.py} & Normalizes pip, npm, uv, bundled, and command-based installation paths. \\
Catalog registries & \path{registry.json}, \path{public_registry.json} & Distribution metadata that lets agents find, install, and invoke interfaces through \hub{}. \\
\bottomrule
\end{tabularx}
\caption{Repository map for terms used throughout the report. Later sections
refer to components by role and use paths only when needed for reproducibility.}
\label{tab:repo-map}
\end{table}

\begin{figure}[t]
\centering
\framedgraphic[0.92\linewidth]{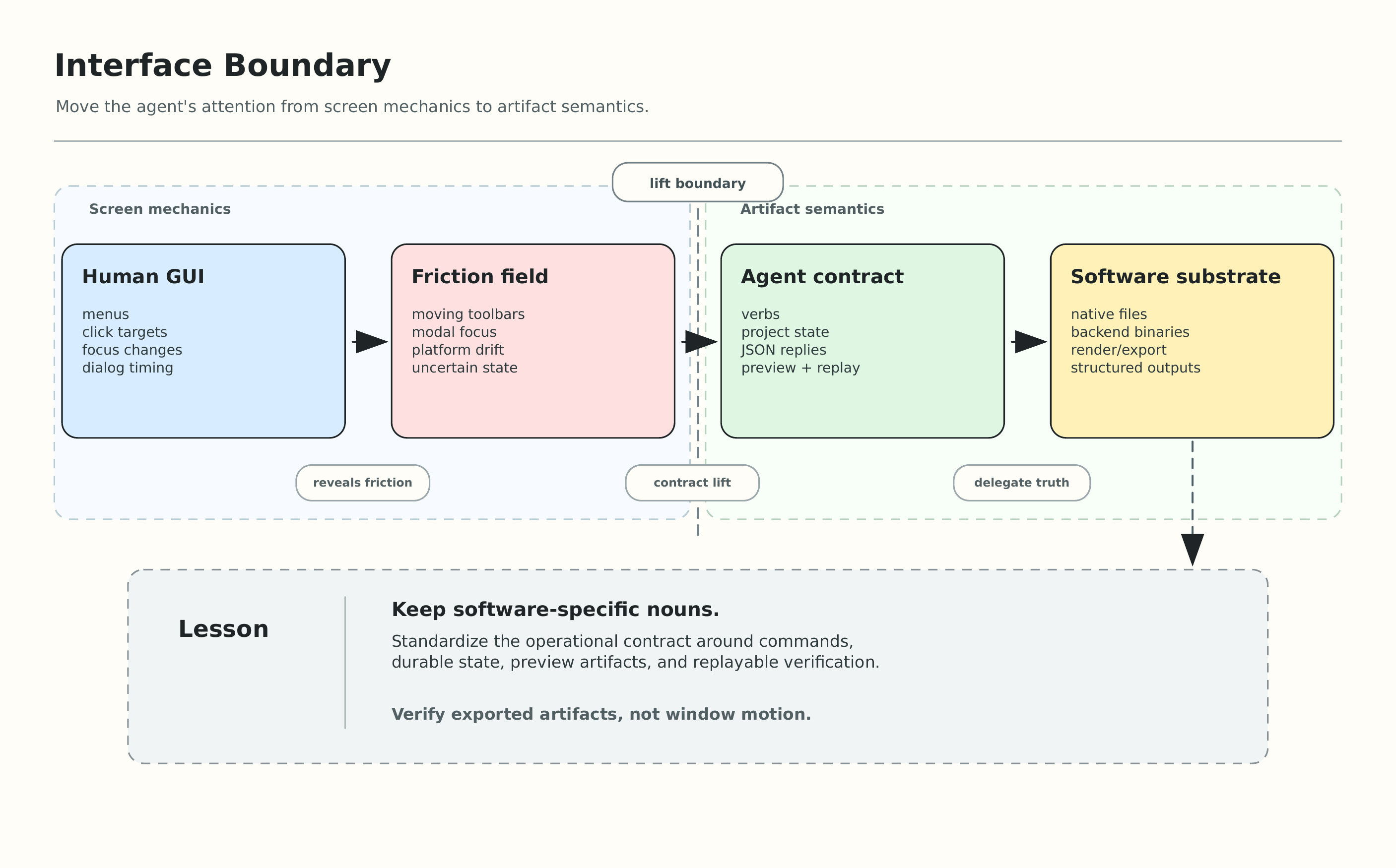}
\caption{The interface boundary. GUI control asks agents to imitate a human
operator. A lifted harness moves the control point toward native state, command
systems, renderers, APIs, and project formats while keeping the real software in
the loop.}
\label{fig:boundary}
\end{figure}

\section{Context: Agentic Models and Agent-Native Interfaces}

Four strands of agent research put pressure on the software interface. Coding
agents assume an executable workspace with files, tests, and package
tools~\cite{qwen-coder-next,swebench,sweagent}. Tool-use training and
API-oriented benchmarks study whether models can select a capability and call it
with the right arguments~\cite{deepseek-v32,toolformer,toolllm,gorilla,apibank}.
Stateful app-world benchmarks add databases, policies, simulated users, and
multi-application workflows~\cite{appworld,taubench}. Multimodal and
computer-use benchmarks keep the visual channel in view: web pages, mobile apps,
desktop apps, and long tool-call workflows. Kimi K2.5 represents the multimodal
model-report side~\cite{kimi-k25}; VisualWebArena, AndroidWorld, OSWorld, and
AppAgent represent benchmark settings where screen state and real applications
are central~\cite{visualwebarena,androidworld,osworld,appagent}.

\system{} belongs to the environment layer required by this line of work. If a
model is trained or evaluated with executable feedback, the environment must expose an
operation boundary. For coding, that boundary is often a repository, a shell, a
test suite, and a package manager~\cite{swebench,sweagent}. For professional
software, the boundary is less standardized: an office suite, a CAD tool, a
graphics application, a video editor, a browser, or a debugger may hide its
useful state behind GUI actions. Early web-agent work exposes this problem in
open-domain and shopping settings~\cite{worldofbits,webshop}. Later web and
enterprise benchmarks make the browser state more realistic and
task-specific~\cite{webarena,mind2web,workarena}.
Visual web and mobile benchmarks show the same issue when pixels and gestures
carry part of the state~\cite{visualwebarena,androidworld,appagent}. OSWorld
extends the setting to arbitrary desktop applications~\cite{osworld}.

The practical question becomes: can we give agents the same style of executable
environment for non-code software that they already have for code? \system{} is
one answer. Each harness turns an application into a stateful tool with a command
tree, a project/session representation, backend invocation, tests, and
agent-readable documentation. The interface is local to the software domain; the
contract is shared with the broader tool-use literature: invoke commands,
inspect state, render or export through the real backend, verify outputs, and
recover from errors. Tool-use work studies command selection and API invocation
itself~\cite{react,toolformer,toolllm}. API-Bank and Gorilla focus more directly
on API-call coverage and correctness~\cite{apibank,gorilla}; reflection work shows
how agents can use feedback from failed attempts~\cite{reflexion}.

\begin{table}[t]
\centering
\small
\begin{tabularx}{\linewidth}{>{\RaggedRight\arraybackslash}p{0.22\linewidth}YY}
\toprule
\tablehead
\textbf{Related-work pattern} & \textbf{Model-side role} & \textbf{\system{} analogue} \\
\midrule
Executable environments & Training and evaluation tasks can be run and checked. & Harness commands mutate real project state and expose inspection commands. \\
Verifiable rewards & Feedback is grounded in tests, tools, or outcome checks. & E2E tests validate native formats, magic bytes, rendered pixels, timelines, and exports. \\
Tool-use trajectories & Agents interleave reasoning and actions over long horizons. & REPL/session state, undo/redo, preview bundles, and trajectory logs persist intermediate work. \\
Agent orchestration & Complex work decomposes into specialized tool calls or subagents. & \hub{} organizes installable harnesses today; CLI Matrix is an experimental direction for scenario-level organization. \\
\bottomrule
\end{tabularx}
\caption{Related agentic systems motivate a software-interface layer. \system{}
supplies explicit environments for real applications and avoids making GUI
pixels the primary execution substrate.}
\label{tab:reference-patterns}
\end{table}

\section{Methodology}

\subsection{The Harness Lift Methodology}

The harness-generation SOP in Table~\ref{tab:repo-map} is the central
methodology. The lifecycle follows the same causal order as the abstract:
discover the application's backend contract, build a stateful harness around
that contract, publish truthful previews from the real backend, generate the
agent-readable skill, and distribute the interface through \hub{}. It begins
with interface archaeology: the builder identifies the backend engine, maps GUI
actions to API calls or command systems, finds the native data model, searches
for existing CLI tools, and studies undo/redo or command patterns. This step
determines whether the harness should manipulate XML, JSON, ODF packages, MLT
timelines, SVG graphs, REST APIs, or a native scripting API. This emphasis on
designing the agent-computer interface follows the same broad lesson as
SWE-agent: interface affordances materially affect agent performance on
interactive software tasks~\cite{sweagent}.

The second phase designs a dual-mode CLI. Every generated harness should work as
both a stateful REPL and a one-shot subcommand tool. It should define command
groups around the domain model of the software: projects, layers, pages, tracks,
clips, filters, parts, scenes, exports, captures, notes, or requests. The common
shape comes from the operational contract: project management, mutation,
inspection, import/export, configuration, session state, and errors that an
agent can understand.

\begin{figure}[t]
\centering
\framedgraphic[0.92\linewidth]{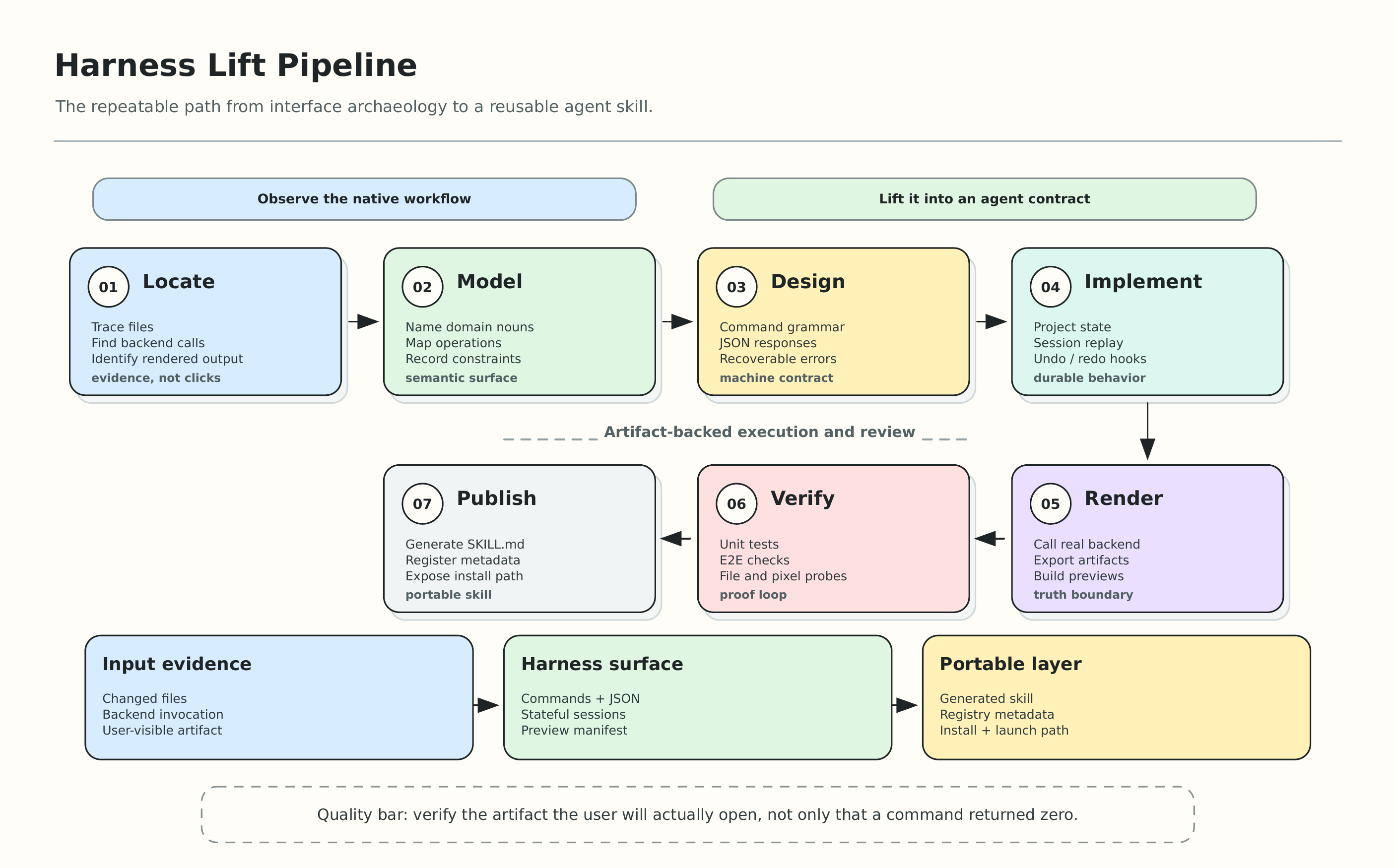}
\caption{The harness lift pipeline. The generated CLI is a new control surface
above native state and backend execution.}
\label{fig:lift}
\end{figure}

The implementation phase then starts from the data layer, adds probes before
mutations, wraps the real backend, and adds rendering/export. The SOP
is explicit that the real software is a hard dependency for rendering and export.
The harness should use tools such as \texttt{libreoffice --headless},
\texttt{blender --background}, \texttt{melt}, \texttt{ffmpeg}, \texttt{inkscape},
\texttt{sox}, native scripting, or service APIs, while avoiding Python
imitations of the application. The GIMP backend execution wrapper, for example,
finds a real GIMP executable, runs Script-Fu in batch mode, and raises an error
when the expected output file is missing.

\begin{codeblock}[Backend export guard]
for name in ("gimp", "gimp-2.10", "gimp-2.99"):
    path = shutil.which(name)
    if path:
        return path
raise RuntimeError("GIMP is not installed...")

cmd = [gimp, "-i", "-b", script, "-b", "(gimp-quit 0)"]
result = subprocess.run(cmd, capture_output=True, text=True)

if not os.path.exists(abs_output):
    raise RuntimeError("GIMP export produced no output file")
\end{codeblock}

The harness also exposes machine-readable output. The standard pattern is a
Click group with \texttt{invoke\_without\_command=True}, a project path, a dry-run
mode, and \jsonflag{} for JSON responses. When invoked without a subcommand, the
tool enters the REPL. This gives agents a stable command entry point and gives
humans a familiar terminal surface. The pattern is deliberately close to
tool-use agent work in which models interleave language reasoning with explicit
actions over an external environment~\cite{react,toolformer}. It also follows
API-oriented work where correct tool selection and argument construction are
measured directly~\cite{toolllm,gorilla,apibank}.

\subsection{A Minimal Agent-Native Software Contract}

A harness can be described as a contract
$H=(S,C,I,R,V,D)$ over a real software backend. $S$ is the persistent state
space: project files, session files, undo history, live preview state, and any
native artifacts needed to reopen the work. $C$ is the command vocabulary:
domain-specific mutations and probes exposed through Click subcommands and the
REPL. $I$ is the inspection surface: JSON status, list, info, schema, history,
and preview-summary commands. $R$ is the rendering/export relation, which
delegates to the real software or its native backend tools. $V$ is the
verification layer: unit tests, E2E tests, subprocess tests, file-format checks,
pixel/media checks, and backend-gated assertions. $D$ is the discovery layer:
\skill{}, registry metadata, install strategy, entry point, and \hub{} records.
The contract therefore joins work on API/tool invocation with work on executable
agent environments. Tool-use work treats commands and APIs as callable actions
with arguments~\cite{react,toolformer,toolllm}. API-centered benchmarks and
systems stress call correctness and API coverage~\cite{gorilla,apibank}. App-world and
user-agent-tool benchmarks add persistent application state and policy
constraints~\cite{appworld,taubench}. Agent-computer-interface work and tool
protocols make the surrounding environment design explicit~\cite{sweagent,mcp}.

This contract is useful because it separates software-specific semantics from
agent-facing invariants. A Blender scene, a Draw.io graph, a LibreOffice
document, and a FreeCAD assembly have different state spaces and command
vocabularies. A useful agent-native interface should still satisfy the same
invariants:

\begin{enumerate}
\item \textbf{Explicit state.} The current workspace must be named, saved, and
inspectable. Hidden GUI focus cannot be the sole source of truth.
\item \textbf{Typed actions.} Commands should map to domain operations, with
mouse gestures kept as fallback implementation details. They should fail with
actionable errors.
\item \textbf{Cheap inspection.} Agents should be able to ask what changed before
committing to the next step.
\item \textbf{Backend truth.} Final render/export should use the real software
or an honest native backend path.
\item \textbf{Programmatic verification.} The harness should check the artifact
that a user will open, including the intermediate file and process return code
only as supporting evidence.
\item \textbf{Discoverability.} An agent should be able to find, install, read,
and invoke the interface from structured metadata.
\end{enumerate}

The contract also clarifies when GUI automation remains necessary. If $S$ is
opaque, $C$ cannot be mapped to native operations, $R$ is unavailable in headless
mode, or $V$ cannot observe the result except through pixels, then a harness may
need accessibility, GUI scripting, or screen control. \system{} is strongest
where the software already contains enough inspectable machinery to make
$H$ meaningful. Its claim has a precise scope: when the backend, file format,
command model, renderer, or API can be lifted, agents should use that lifted
contract first. Earlier web-agent work studied open-domain and shopping
settings~\cite{worldofbits,webshop}; later web and enterprise benchmarks add
realistic sites and knowledge-work workflows~\cite{webarena,mind2web,workarena}.
Visual and mobile-agent work remains important where pixels, gestures, and app
screens are the available control surface~\cite{visualwebarena,androidworld,
aitw,appagent}. Desktop benchmarks such as
OSWorld cover applications where arbitrary GUI state still matters~\cite{osworld}.
Earlier GUI automation systems showed that screenshot search and pixel-based
interface recovery can be useful when structure is unavailable~\cite{sikuli,
prefab}.

\subsection{Verification Becomes Execution}

Many GUI computer-use tasks create an artificial asymmetry. The final artifact is
structured and easy to inspect, while the construction path is framed as pixel
observation and clicking. A spreadsheet can be parsed. A slide deck is a package. A
diagram is XML. A video editor timeline has tracks and timecodes. A CAD project
has geometry, constraints, and exportable views. If the artifact can be inspected
programmatically, then in many cases it can also be constructed programmatically.
This is the software-artifact analogue of the move from static benchmark answers
to executable environments in coding, web, and desktop-agent
evaluation. Coding benchmarks use repositories and tests as
validators~\cite{swebench,sweagent}; web and enterprise benchmarks use browser or service
state~\cite{webarena,mind2web,workarena}; desktop benchmarks use OS-level setup
and execution-based checks~\cite{osworld}.
\system{} extends the same argument to artifact-producing desktop software:
when the success condition is a file, render, database state, or inspectable
project graph, the construction interface should expose that state directly
instead of forcing every step through visual manipulation.

\system{} restores symmetry by making the construction and verification surfaces
meet at the same artifact boundary. A Draw.io harness can create shapes and
connectors, then tests can parse the resulting \texttt{mxfile}. A LibreOffice
harness can construct ODF state, export through headless LibreOffice, and verify
PDF or OOXML output. A video harness can mutate MLT state, render through the
real backend, and verify duration, frames, or media properties. This is why the
testing rules in the SOP are strict: unit tests and process exit codes provide
incomplete evidence on their own.

\begin{figure}[t]
\centering
\framedgraphic[0.92\linewidth]{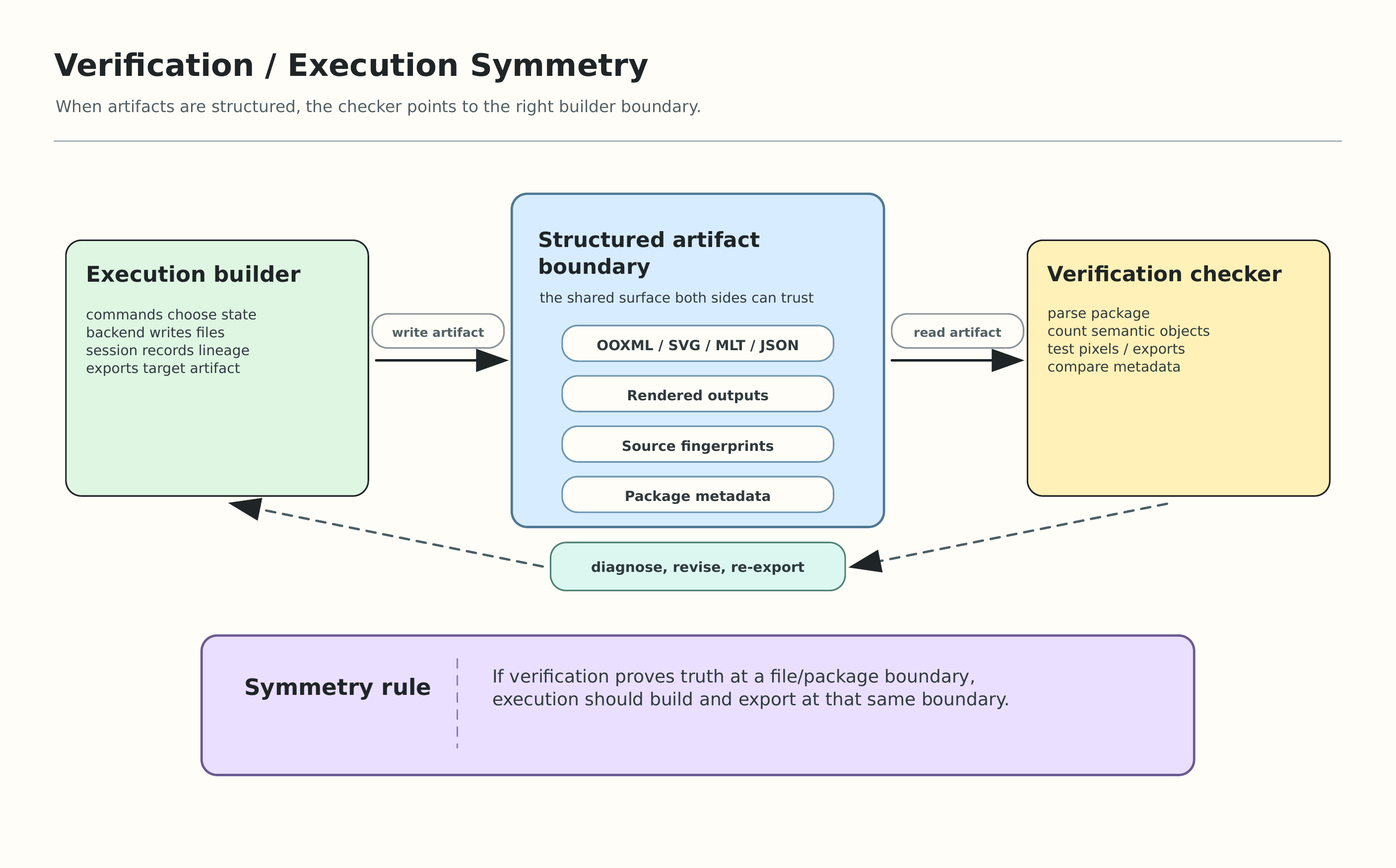}
\caption{Verification/execution symmetry. When commands and checks share the
same artifact boundary, agents can build, inspect, render, verify, and replay
without switching to a pixel-only mental model.}
\label{fig:symmetry}
\end{figure}

The repository encodes this principle in test requirements. Harnesses should
test core functions with synthetic data, then run E2E tests that generate real
files, call real software, and verify the output that a user would open. The SOP
specifically calls for magic-byte checks, ZIP structure validation for OOXML,
pixel-level analysis for video/images, audio analysis, duration checks, and CLI
subprocess tests against installed commands. This makes tests part of interface
design. A weak harness only says ``the command returned zero.'' A stronger
harness says ``the command generated the project, the backend rendered it, and
the output has the expected structure or observable content.''

The same lesson appears in related work on agentic coding and tool use:
Qwen3-Coder-Next uses executable
coding tasks because coding-agent behavior needs executable
feedback~\cite{qwen-coder-next}.
DeepSeek-V3.2 synthesizes tool environments because robust agent behavior needs
interactive feedback~\cite{deepseek-v32}. \system{} applies the same logic to
real applications: artifacts should be executable, checkable, and visible through
a trustworthy output path.

\subsection{State Is the Agent Workspace}

Stateless shell commands help with isolated calls. Serious agent workflows need
more continuity. An agent may need to open a project, add objects, inspect the state, preview a
change, undo a failed attempt, revise parameters, export, and cite the final
artifact. A human GUI carries continuity through windows, selections, recent
files, focus, visual memory, and unsaved state. A command-line agent needs the
same continuity represented as files, history, and commands.
This requirement is consistent with prior agent work that treats state,
environment feedback, memory, and executable skills as central to long-horizon
behavior~\cite{react,reflexion,voyager}.

\system{} harnesses therefore make state a first-class surface. JSON project
files give agents an inspectable workspace. Session files record current project
paths and modification state. Undo/redo stacks make exploration recoverable.
Dry-run modes help agents test plans before mutation. Deterministic JSON makes
diffing useful. File locking prevents concurrent writes from corrupting a
session. The FreeCAD session implementation is a representative example: it
saves undo snapshots, maintains redo history, tracks modification state, and
persists JSON under an exclusive lock when the platform supports it.

\begin{codeblock}[Session undo snapshot]
entry = {
    "timestamp": time.time(),
    "description": description,
    "state": copy.deepcopy(self.project),
}
self._undo_stack.append(entry)
self._redo_stack.clear()
self._modified = True
\end{codeblock}

This state model gives the CLI the qualities of a desktop session without
requiring the desktop. It also changes how agents can recover. A screenshot
leaves the agent inferring whether a change succeeded; the harness lets it ask
for \texttt{session status --json}, inspect the project graph, list history,
undo, or render a preview bundle. The workspace for agents becomes a state graph
with executable transitions and inspectable history.
The difference from a plain command wrapper is that the CLI itself becomes an
agent workspace with memory, recovery, and feedback loops, far beyond a thin
process invocation surface~\cite{reflexion,voyager,sweagent}.

\subsection{Preview as the Display Protocol}

Agents still need visual feedback. Humans still need to inspect outputs. The
open question is the contract that carries intermediate rendered state.
Multimodal GUI agents and desktop benchmarks show why visual
state remains important and why it needs reliable task context plus
execution-grounded evaluation. VisualWebArena isolates visually grounded web
tasks~\cite{visualwebarena}; AndroidWorld and Android in the Wild expose mobile
device-control settings~\cite{androidworld,aitw}; OSWorld studies arbitrary
desktop applications~\cite{osworld}; and AppAgent studies smartphone-app use
through multimodal interaction~\cite{appagent}. \system{}
separates preview from GUI streaming. The preview
protocol defines a stable on-disk bundle with \texttt{manifest.json},
\texttt{summary.json}, and an \texttt{artifacts/} directory containing images,
clips, JSON dumps, models, or other inspection artifacts. Live workflows add
\texttt{session.json} for the mutable current head and \texttt{trajectory.json}
for append-only command-to-preview history.

The producer/consumer split is crucial. A software harness publishes preview
state through \texttt{cli-anything-\textless{}software\textgreater{} preview ...}. It owns the real
backend, source fingerprinting, recipes, and truthful artifact generation.
\hub{} consumes already-published state with \texttt{cli-hub previews inspect},
\texttt{html}, \texttt{watch}, and \texttt{open}. It reads preview bundles
without rendering or mutating the project. This keeps domain truth inside the harness and keeps viewing generic
across CAD, video, 3D, GPU captures, and diagrams.

\begin{figure}[t]
\centering
\framedgraphic[0.92\linewidth]{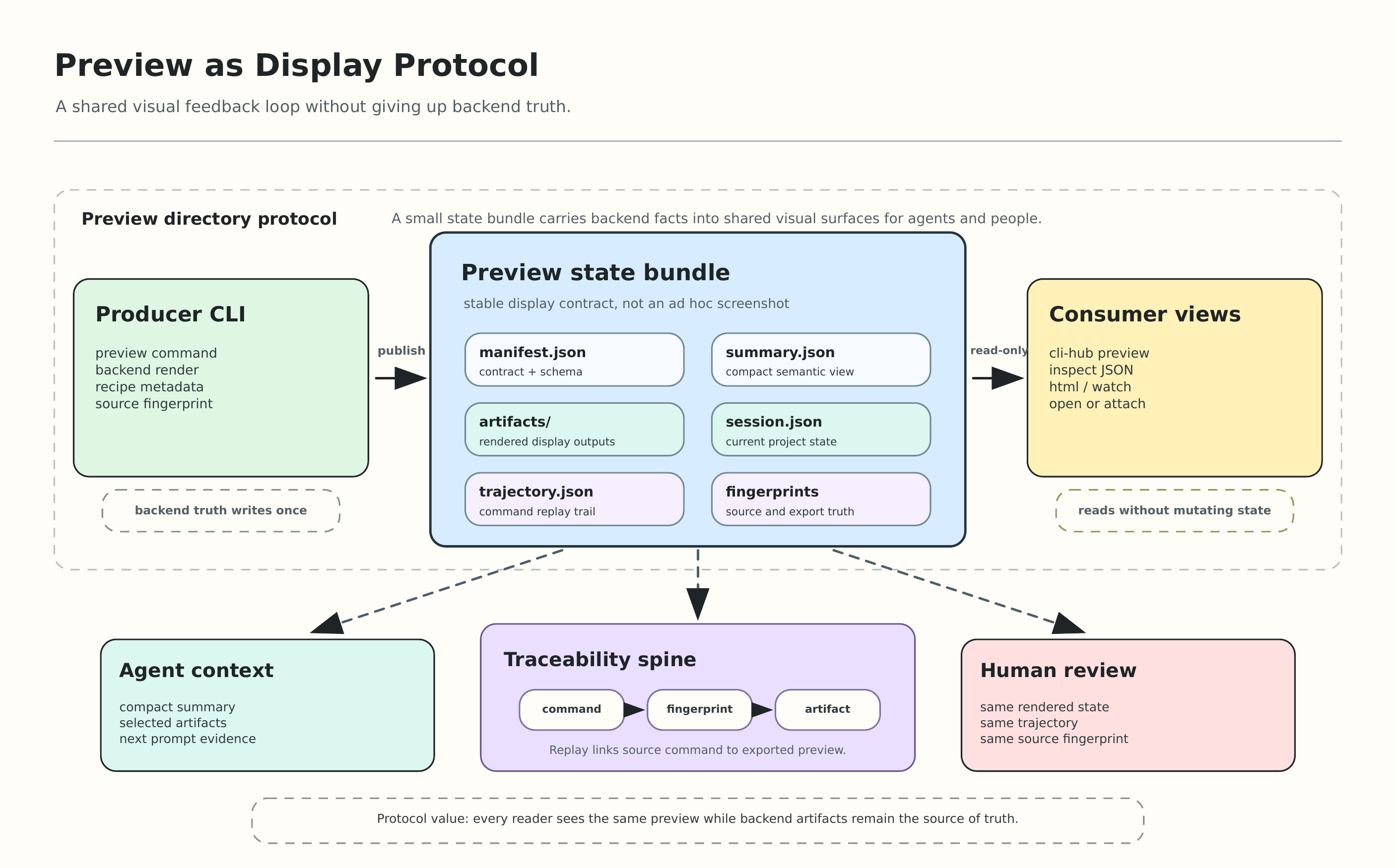}
\caption{Preview as display protocol. Harnesses publish preview bundles and
live-session state; \hub{} reads that state as a generic viewer.}
\label{fig:preview}
\end{figure}

The current implementation includes \MetricPreviewHelpers{} preview-capable
helper integrations: Shotcut, Openscreen, Blender, FreeCAD, and RenderDoc.
FreeCAD and Blender are the strongest demonstrations because they
produce trajectories where each command is tied to a real preview state. The
same mechanism serves agents as much as human demos: mutate the project, publish
a preview, inspect a compact summary or rendered artifact, and continue from a
known state.

Truthfulness is the central design constraint. The protocol explicitly rejects
toy renderers and GUI screen scraping as a source of preview truth. A preview
should be a cheap intermediate view of the real project or capture state. If a
helper camera, light, or temporary rig is inserted for visualization, the bundle
should mark and explain that limitation. Trustworthy preview quality matters
most. In this sense, preview is an inspectable display protocol
for executable environments that continues to rely on the underlying software's
renderer or state model. This keeps the visual evidence needed by multimodal
web and mobile agents~\cite{visualwebarena,androidworld,appagent}, while
preserving execution-grounded desktop state as emphasized by OSWorld~\cite{osworld}.

\subsection{Discovery, Distribution, and \hub{}}

A CLI harness is incomplete if an agent cannot discover it, install it, read its
capabilities, and recover from installation or backend errors. This is why
\system{} treats \skill{} and \hub{} as interface infrastructure alongside the
CLI surface. The current system includes \MetricSkillFiles{} companion skills for
CLIs. These skills summarize command groups, examples, JSON usage, prerequisite
software, and agent-specific constraints, giving agents task-time capability
descriptions instead of forcing them to infer behavior from help text alone.
This discovery layer follows the
same pressure identified by tool-use work: agents need a reliable way to select
and invoke external capabilities, with prompt-time descriptions serving as one
piece of the runtime surface. Tool-use and API-call work motivates accurate tool
selection and invocation~\cite{toolformer,toolllm}. API-centered systems and
benchmarks stress call correctness and coverage~\cite{gorilla,apibank}; the
Model Context Protocol (MCP) motivates a protocol boundary for exposing tools and
context~\cite{mcp}.

\hub{} adds the package and registry layer. The local registry contains
\MetricHarnessEntries{} \system{} CLI entries across \MetricHarnessCategories{}
categories. The public registry contains \MetricPublicEntries{} third-party
entries across \MetricPublicCategories{} categories. The installer dispatches across pip,
npm, uv, bundled/system tools, and command-based installers. Listing and search
commands support JSON output, and preview commands provide a generic consumer for
preview bundles and live sessions.

\begin{table}[!tbp]
\centering
\small
\begin{tabularx}{\linewidth}{>{\RaggedRight\arraybackslash}p{0.32\linewidth}>{\RaggedLeft\arraybackslash}p{0.14\linewidth}Y}
\toprule
\tablehead
\textbf{Ecosystem coverage signal} & \textbf{Current value} & \textbf{Why it matters} \\
\midrule
Harness catalog entries & \MetricHarnessEntries{} & Agent-facing interfaces for generated and contributed harnesses. \\
Public third-party CLI entries & \MetricPublicEntries{} & External CLIs normalized under the same discovery/install layer. \\
Companion skills for CLIs & \MetricSkillFiles{} & Machine-readable usage surfaces for agents and humans. \\
Harness catalog categories & \MetricHarnessCategories{} & Capability browsing beyond package names. \\
Public catalog categories & \MetricPublicCategories{} & Mixed install strategies and external tool coverage. \\
Preview-capable integrations & \MetricPreviewHelpers{} & Early cross-harness display protocol adoption. \\
\bottomrule
\end{tabularx}
\caption{Current production snapshot for catalog and discovery coverage.}
\label{tab:counts}
\end{table}

The current \hub{} is still partly human-facing. The next agent-first version
should expose structured capability search, install plans, status diagnostics,
direct skill fetch, preview metadata, and a stable schema that agents can query
during work. Protocol work such as MCP points in the same direction for
tool/context interoperability; \hub{} is the package, skill, and capability
layer for CLI harnesses rather than a model-provider protocol~\cite{mcp}. At
that point discovery becomes part of runtime: the agent asks for
a capability, receives candidate interfaces, installs or locates one, loads the
skill, checks backend availability, and continues the task.

\begin{figure}[H]
\centering
\framedgraphic[0.92\linewidth]{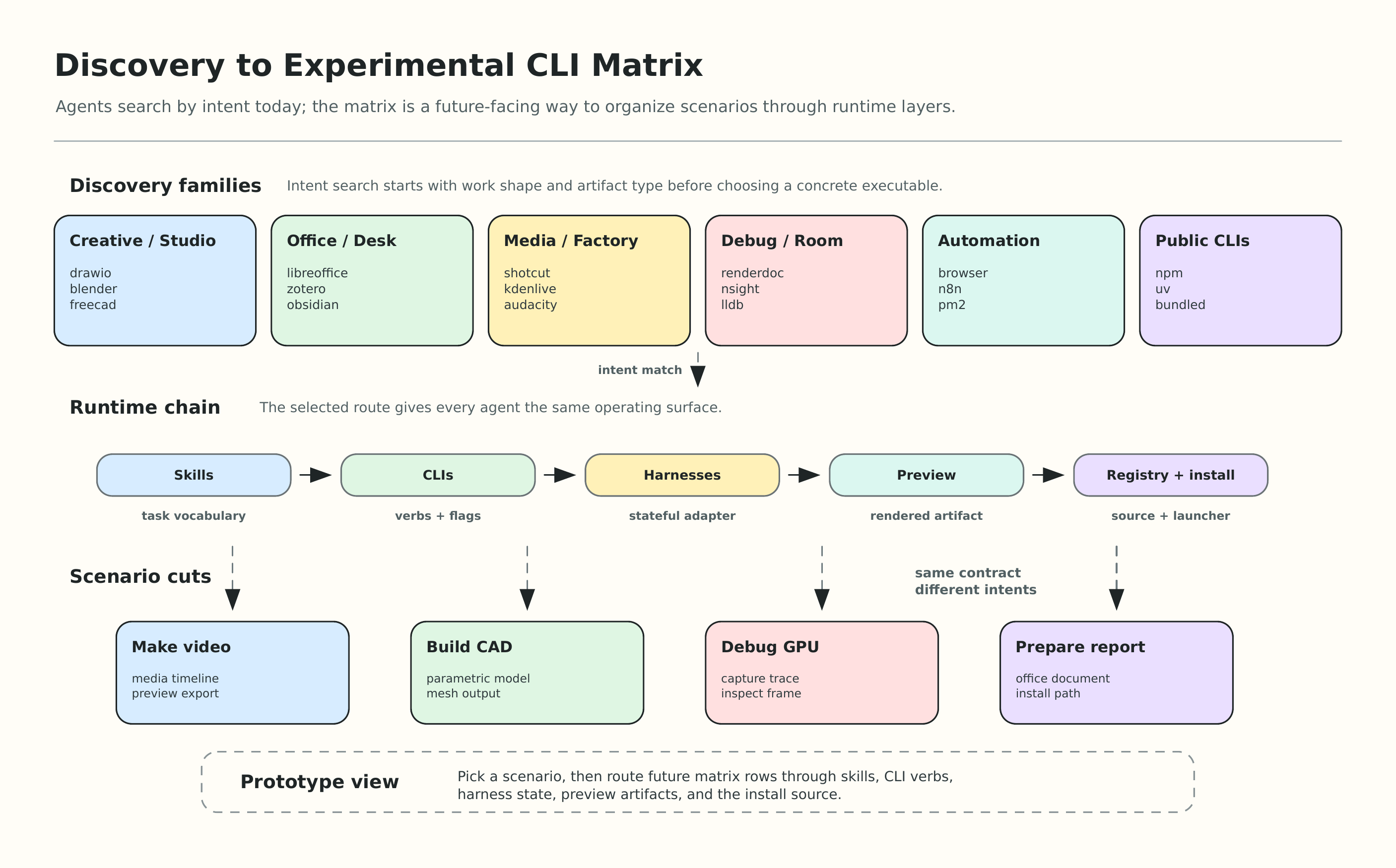}
\caption{Discovery toward an experimental CLI Matrix. Once many interfaces
exist, agents will need to choose tools by scenario, capability, and package
metadata together.}
\label{fig:matrix}
\end{figure}

\section{Case Study}

\subsection{Growing Blender Into an Agent-Native CLI}

Blender is a useful stress test for \system{} because it spans full scene
construction across a large application surface. A useful Blender interface has to represent
scene-level settings, mesh primitives, transforms, materials, modifier stacks,
cameras, lights, animation, render configuration, backend execution, and visual
feedback. The current Blender harness therefore shows how a real, complicated
application grows an agent-facing command surface without replacing the real
renderer.

The current Click tree contains \MetricBlenderGroups{} group nodes including the
root command group and nested live-preview group. It exposes
\MetricBlenderCommands{} command functions: \MetricBlenderPublicCommands{}
public commands and \MetricBlenderHiddenCommands{} hidden monitor command used
by live preview. The surface follows the backend concepts an agent needs to
construct and check a scene:
\texttt{scene}, \texttt{object}, \texttt{material}, \texttt{modifier},
\texttt{camera}, \texttt{light}, \texttt{animation}, \texttt{render},
\texttt{preview}, and \texttt{session}.

\begin{figure}[H]
\centering
\framedgraphic[0.94\linewidth]{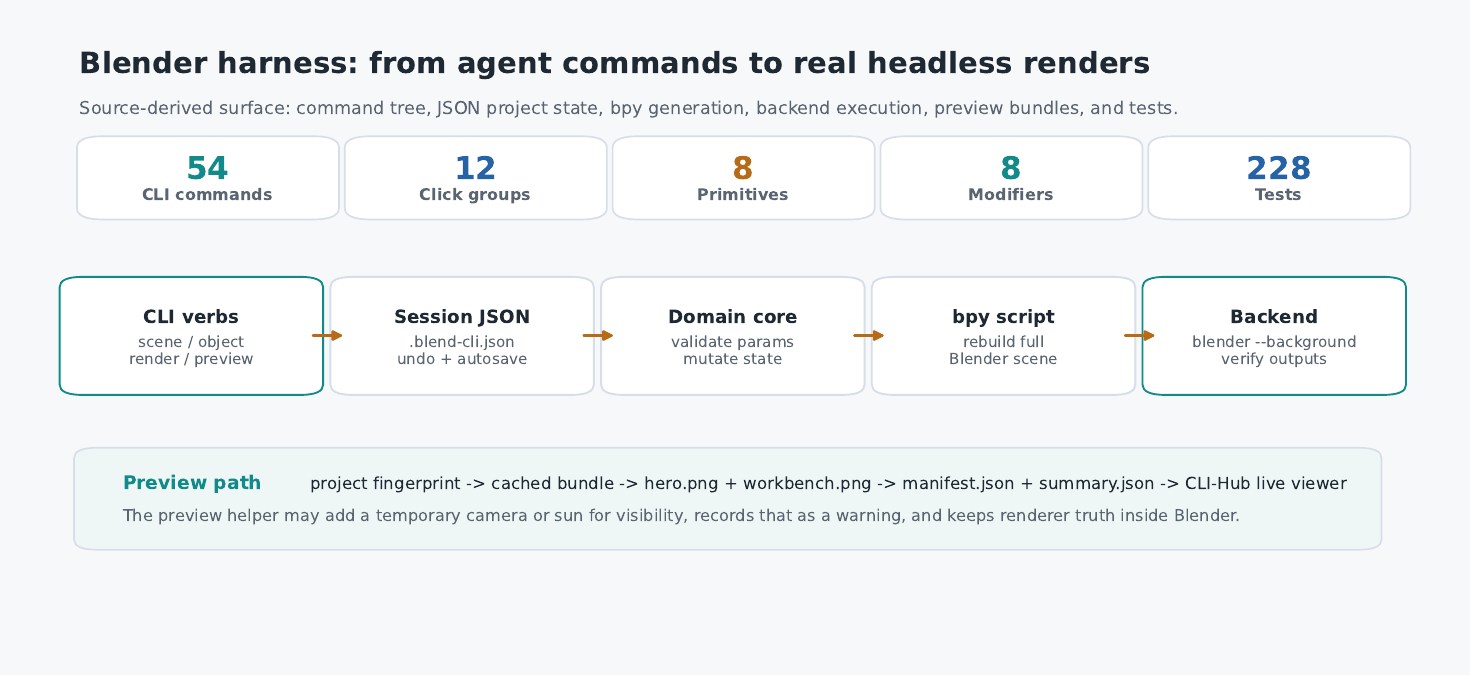}
\caption{Blender case-study pipeline. Commands mutate a JSON scene contract,
domain modules validate the state, the bpy generator rebuilds the scene, and the
real Blender backend renders or publishes preview artifacts.}
\label{fig:blender-pipeline}
\end{figure}

\subsubsection{From Blender's Native Architecture to Harness Growth}

The first design question is where Blender already has a stable, executable
boundary. Natively, Blender is a GUI
wrapped around a large scene database and execution engine. A \texttt{.blend}
file stores data-blocks such as scenes, objects, meshes, materials, cameras,
lights, animation curves, render settings, and collections. The 3D viewport,
timeline, properties editor, and outliner are human-facing views and editors
over that graph. Under those views sits Blender's Python API,
\texttt{bpy}, which exposes the same object graph, operators, RNA properties,
and render configuration that the GUI manipulates. Below that are the render
engines--Cycles, EEVEE, and Workbench--which turn the scene graph into images,
animation frames, and preview artifacts.

\begin{figure}[!tbp]
\centering
\framedgraphic[0.94\linewidth]{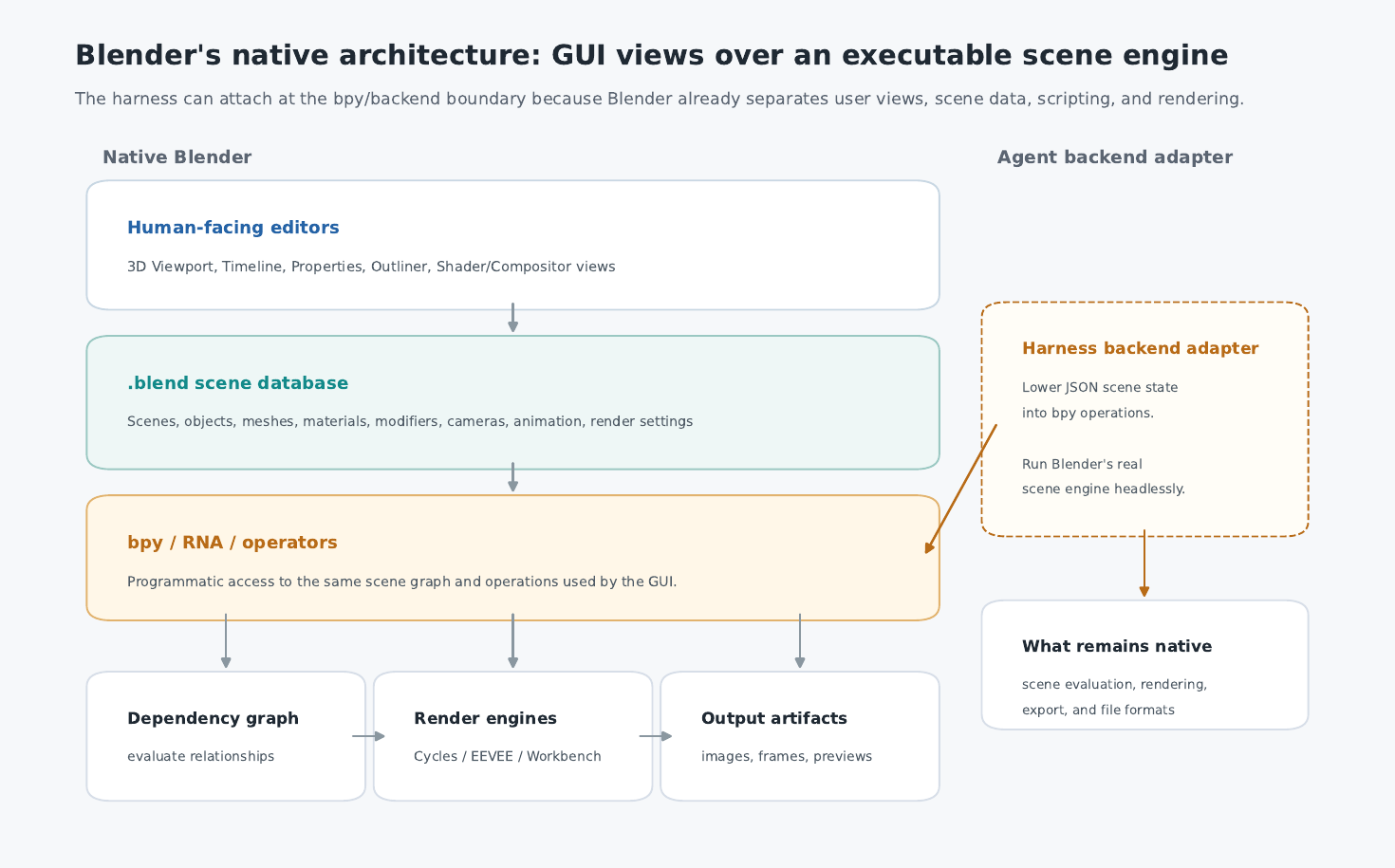}
\caption{Blender's original software architecture. The GUI editors are views
over a \texttt{.blend} scene database; \texttt{bpy}, RNA, and operators expose
that same state programmatically; the dependency graph and render engines
produce output artifacts. CLI-Anything attaches above the backend scripting and
rendering layers, below the GUI event machinery.}
\label{fig:blender-original-architecture}
\end{figure}

This architecture gives \system{} a natural injection point. The harness can
avoid GUI clicking and renderer implementation because Blender already exposes a
backend scripting layer through \texttt{bpy}, RNA
properties, operators, dependency-graph evaluation, and native render engines.
The harness layer can therefore sit outside Blender as a stateful command
surface, generate a \texttt{bpy} program, and ask the real Blender process to
execute it without replaying GUI events. The injected layer is thin in the right
place: it translates agent intentions into a stable scene contract and into
native Blender Python, while Blender remains responsible for scene construction
semantics, version-specific behavior, dependency evaluation, file formats, and
final rendering.

This is why the case study centers on architecture and backend reuse. \system{}
uses Blender's own backend layers as the source of truth. A precise scene
contract, a correct \texttt{bpy} lowering pass, and a backend wrapper are enough
to invoke Blender's scene engine and check the resulting artifact without
privileged access to Blender internals.

The remaining design problem is the project format. Blender's native
\texttt{.blend} files are binary and excellent for Blender itself. They are a
poor primary workspace for an agent that needs diffs, inspection,
undo/redo, and JSON responses after every command. The harness therefore keeps
an agent-facing \texttt{.blend-cli.json} contract as the working state and treats
\texttt{.blend} and rendered images as backend artifacts. This creates a
four-stage mapping:
\begin{enumerate}
\item CLI commands mutate a JSON scene contract that an agent can inspect.
\item Domain modules validate Blender concepts before they become backend code.
\item The scene-contract lowering pass turns the JSON graph into a complete
\texttt{bpy} script.
\item The backend execution wrapper runs the real Blender binary and verifies
that the expected artifact exists.
\end{enumerate}

The current backend design follows directly from that mapping.
The scene model creates a \texttt{.blend-cli.json} contract with
\texttt{scene}, \texttt{render}, \texttt{world}, \texttt{objects},
\texttt{materials}, \texttt{cameras}, \texttt{lights}, \texttt{collections},
and \texttt{metadata}. The session layer wraps that contract in an agent
workspace: project path, modified state, undo/redo snapshots, history, and
autosave behavior. This lets an agent inspect the graph, recover from mistakes,
and continue across command turns without depending on a window being open.

The command tree grows by preserving Blender's semantic structure. Object and
material components map primitive creation, transforms, material creation, and
assignment into checked JSON operations. The modifier component defines
\MetricBlenderModifiers{} modifier types with typed parameters and ranges,
mirroring Blender's modifier-stack idea in a form an agent can inspect. The
lighting, animation, and render components map cameras, lights, frame ranges,
keyframes, FPS, interpolation, render engines, presets, resolution, samples, and
output formats into validated scene state. The command entry point is therefore
a surface over a typed Blender scene model with coherent backend semantics.

The render path is where the harness meets the real application.
The scene-contract lowering pass translates the JSON scene into a full Blender
Python script: clear the default scene, apply unit and frame settings, set
render and world state, create materials, instantiate primitives, restore
parenting, attach modifiers, create cameras and lights, insert keyframes, and
call \texttt{bpy.ops.render.render}. The backend execution wrapper then locates
the installed \texttt{blender} binary, runs the generated script through
Blender's headless execution path, checks the return code, verifies that an
output artifact was actually produced, resolves Blender's frame-number suffixes,
and returns the method, format, Blender version, and file size. The
generated EEVEE script path is version-tolerant: newer Blender builds can select
\texttt{BLENDER\_EEVEE\_NEXT}, while Blender 4.0 still selects
\texttt{BLENDER\_EEVEE}. The principle is the same as the rest of \system{}:
the harness should absorb backend variability and expose stable agent-facing
verbs.

\begin{table}[!tbp]
\centering
\small
\begin{tabularx}{\linewidth}{>{\RaggedRight\arraybackslash}p{0.22\linewidth}YY}
\toprule
\tablehead
\textbf{Native Blender layer} & \textbf{\system{} implementation} & \textbf{Agent-facing result} \\
\midrule
Scene database and binary \texttt{.blend} state &
Scene model and session manager &
Create, open, save, inspect, undo, redo, and persist an inspectable \texttt{.blend-cli.json} workspace. \\
Objects, meshes, materials, modifiers, cameras, and lights &
Object, material, modifier, and lighting components &
Build Blender scene resources with checked parameters, stable IDs, and JSON output after mutations. \\
Timeline, render settings, and \texttt{bpy} execution model &
Animation and render components plus the scene-contract lowering pass &
Set frame ranges, keyframes, render presets, output formats, and lower the scene contract into executable \texttt{bpy} scripts. \\
Background Blender process and render engines &
Backend execution wrapper &
Run the real Blender binary headlessly and reject missing or empty render outputs. \\
Viewport feedback and render-backed preview artifacts &
Preview publisher and bundle writer &
Publish preview bundles, live-session metadata, trajectory history, and viewer commands for \hub{}. \\
Validation evidence for the lifted interface &
Unit tests and backend-gated artifact checks &
Check the scene contract in isolation, then verify generated scripts, native execution, preview bundles, and rendered outputs through Blender. \\
\bottomrule
\end{tabularx}
\caption{How the Blender harness turns backend concepts into a comprehensive
CLI surface. The command tree grows from stateful, testable backend components
and domain concepts.}
\label{tab:blender-case-study}
\end{table}

A representative workflow illustrates the contract. The agent creates or opens
a scene, issues domain commands that mutate JSON state, asks for structured
inspection, then renders or previews through Blender:

\begin{codeblock}[Blender harness workflow]
cli-anything-blender scene new --profile product_render --output scene.blend-cli.json
cli-anything-blender --project scene.blend-cli.json \
  object add cube --name Body --location 0,0,0
cli-anything-blender --project scene.blend-cli.json \
  material create --name BrushedMetal --metallic 0.8
cli-anything-blender --project scene.blend-cli.json material assign 0 0
cli-anything-blender --project scene.blend-cli.json \
  modifier add bevel --object 0 --param width=0.08 --param segments=3
cli-anything-blender --project scene.blend-cli.json camera add --name MainCam --active
cli-anything-blender --project scene.blend-cli.json light add sun --name KeySun
cli-anything-blender --project scene.blend-cli.json \
  preview capture --recipe quick --force
\end{codeblock}

\subsubsection{Preview and Test Evidence}

Preview is implemented as a first-class Blender pathway. The preview publisher
currently exposes \MetricBlenderPreviewRecipes{} preview recipe,
\texttt{quick}. It fingerprints the project state, prepares or reuses a
cacheable bundle, ensures the scene is viewable by adding a preview camera or
sun only when needed, renders both an Eevee-style hero image and a Workbench
structure image, and writes
\texttt{manifest.json}, \texttt{summary.json}, and artifact records. Live
preview adds \texttt{session.json}, a \texttt{current} link,
\texttt{trajectory.json}, poller metadata, and \hub{} watch/inspect/html
commands. The important design point is that the preview stays cheap and
inspectable while still being produced through Blender's renderer and recorded
as a bundle an agent can reason about.

The test suite mirrors this layered design. Unit tests cover scene creation,
object operations, materials, modifiers, lighting, animation, render settings,
session history, and preview bundle bookkeeping using synthetic state.
E2E tests cover scene roundtrips, generated bpy script validity, realistic
workflows, CLI subprocess behavior, preview capture and live refresh, backend
discovery, real headless renders, PNG magic-byte checks for preview artifacts,
and direct execution of minimal bpy scripts through Blender. The important
evidence is the validation boundary: local scene logic is checked in isolation,
and final artifacts are checked through the native backend wherever possible.
For a lifted creative tool, a command succeeds only when the produced scene,
preview bundle, or render artifact survives the same backend path a user would
eventually trust.

\subsection{Slay the Spire II: When the Backend Has to Be Discovered Inside the Running App}

The Blender case works because the application already gives us a mature backend
surface. \texttt{bpy} exposes the scene graph, operators, render configuration,
and artifact production path. Slay the Spire II is a different kind of case. The
useful state lives inside a running game process. The state is temporal, screen
dependent, and split across game systems and UI nodes: run state, combat state,
current room, map screen, reward overlays, card-selection screens, merchant
screens, menu buttons, and transient hand-selection prompts. A CLI for this
software starts with runtime archaeology: finding where decisions enter the game,
which objects carry authoritative state, and which operations can be made stable
enough for an external agent.

The published \texttt{slay\_the\_spire\_ii} harness takes that path. Its Python
side exposes \MetricStsTwoCliCommands{} Click commands, while the in-game bridge
normalizes the game into \MetricStsTwoDecisionStates{} decision states and
\MetricStsTwoBridgeActions{} single-player action verbs. The bridge is not only
an input shim; it creates the backend boundary that the original game does not
expose. This is the important engineering contrast with Blender. In Blender, the
harness finds a backend boundary that already exists. In Slay the Spire II, the
harness first creates a small backend boundary from inside the application.

This also exposes a training problem that is easy to miss when a game is treated
only as a black-box GUI. Keyboard-and-mouse control is a low-level motor
interface, not a semantic action interface. The same intent, such as playing one
card on one monster, can be realized through many different traces: different
cursor paths, hover states, click timings, drag gestures, screen coordinates,
animation delays, and UI resolutions. Those traces may all produce the same game
transition, but they look different to a learner. A policy trained on them has to
spend capacity modeling accidental input variation instead of the game decision
itself. A CLI action collapses that equivalence class into one stable command
with explicit arguments. \texttt{play-card} means playing a card; the bridge, not
the model, owns the unstable details of how that action is delivered to the
runtime.

\begin{figure}[!tbp]
\centering
\framedgraphic[0.94\linewidth]{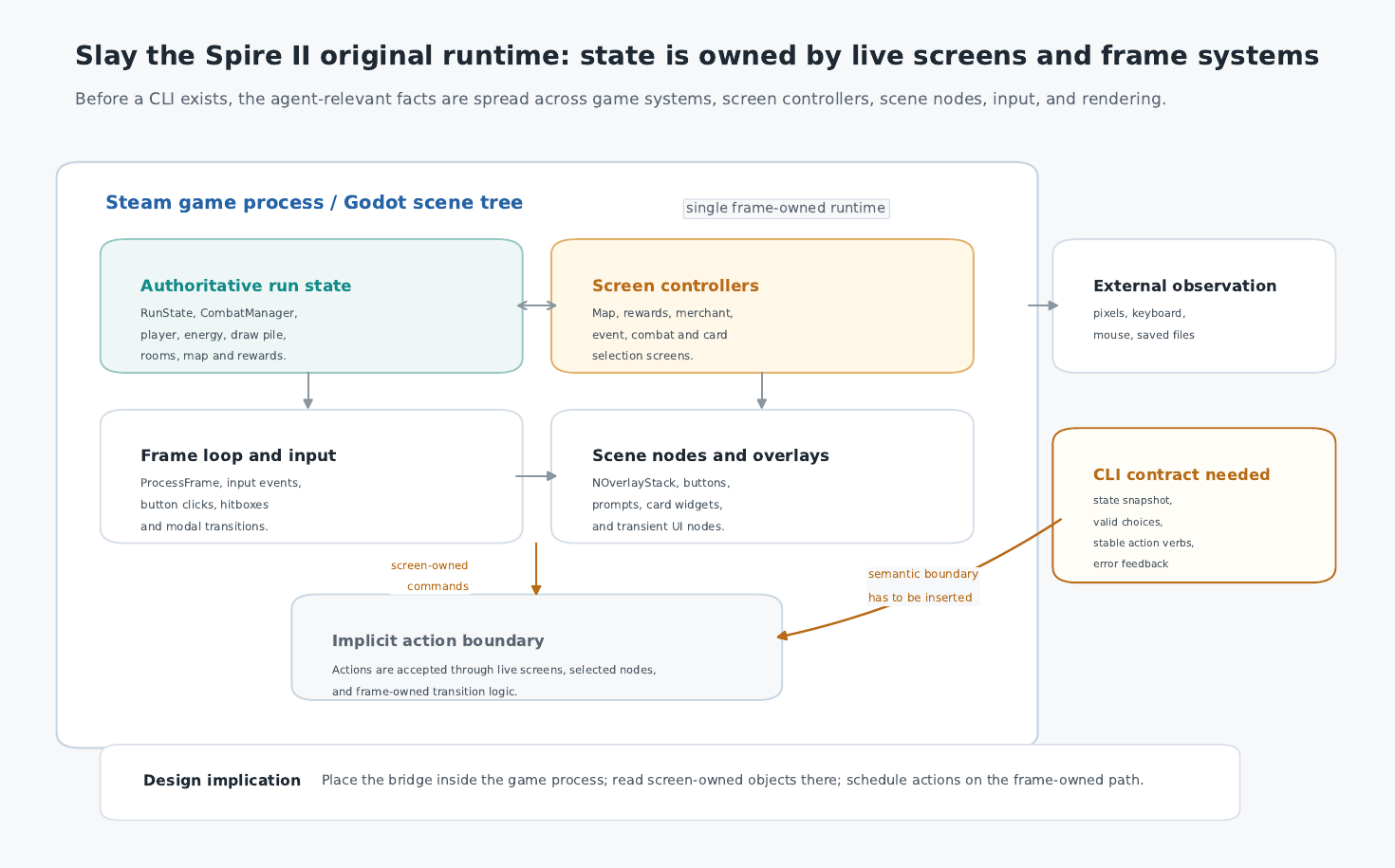}
\caption{Slay the Spire II before the harness. Run state, combat state, room
state, overlays, screen nodes, input handlers, and rendering share one live game
runtime. The useful control boundary sits inside the process, beside the objects
that own the current decision.}
\label{fig:sts2-original-runtime}
\end{figure}

\begin{figure}[!tbp]
\centering
\framedgraphic[0.94\linewidth]{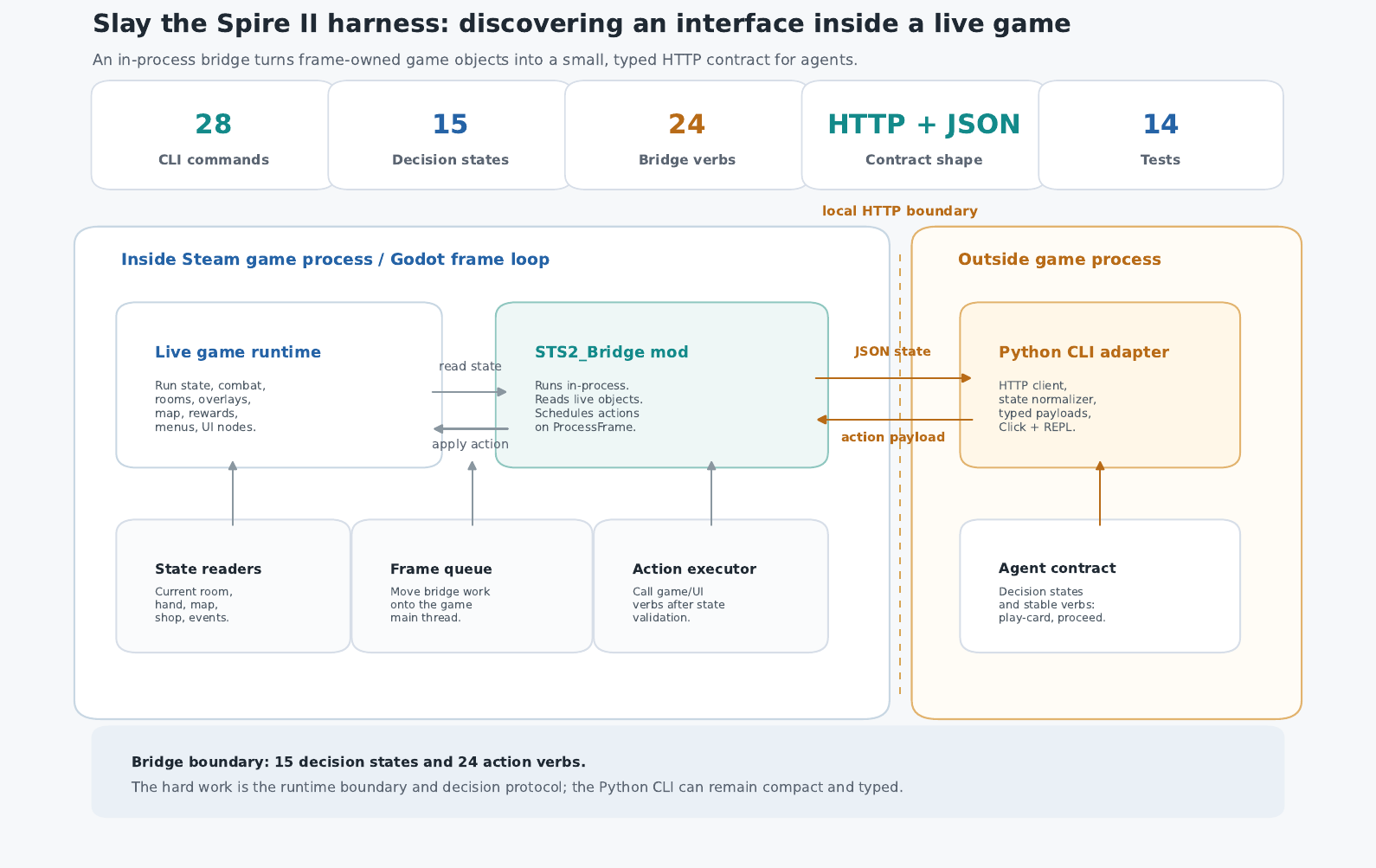}
\caption{Slay the Spire II in-process bridge. The CLI speaks HTTP from outside
the game, while \texttt{STS2\_Bridge} runs inside the Godot/MegaCrit runtime,
schedules work on the main thread, reads live screens and run state, and exposes
stable decision/action contracts.}
\label{fig:sts2-bridge}
\end{figure}

\subsubsection{The Bridge Has to Live Where the State Lives}

Process placement follows a simple rule: put the harness boundary next to the
objects whose names matter. From outside the process, a turn in a deckbuilder is
mostly a rendered screen. From inside the process, the same turn is a structured
state: player energy, hand cards, draw pile, monster intents, current room,
reward list, map choices, event options, shop inventory, and the overlay currently
blocking progress. The bridge lives inside the game because those names already
exist there.

This changes the meaning of ``CLI wrapping.'' The wrapper is a two-part system.
The C\# bridge is the privileged reader and actor inside the runtime. The Python
CLI is the small external language that agents call. Keeping those roles separate
gives the harness a clean failure boundary: the mod owns Godot and MegaCrit
objects; the CLI owns parsing, JSON output, timeouts, REPL behavior, and
agent-facing command names.

The bridge therefore starts by attaching to the Godot main loop and opening a
loopback HTTP listener:

\begin{codeblock}[STS2 bridge initialization]
var tree = (SceneTree)Engine.GetMainLoop();
tree.Connect(SceneTree.SignalName.ProcessFrame, Callable.From(ProcessMainThreadQueue));

_listener = new HttpListener();
_listener.Prefixes.Add("http://localhost:15526/");
_listener.Prefixes.Add("http://127.0.0.1:15526/");
_listener.Start();

_serverThread = new Thread(ServerLoop)
{
    IsBackground = true,
    Name = "STS2_Bridge_Server"
};
_serverThread.Start();
\end{codeblock}

The local HTTP boundary is deliberately ordinary. Python can call it through a
small client. A human can inspect it with \texttt{curl}. The bridge can return
plain JSON or markdown without teaching the game about terminals, Click, or
agent sessions. The endpoint surface stays tiny: \texttt{GET} reads the current
state, and \texttt{POST} sends one action payload. The hard engineering remains
inside the mod, where the runtime objects are available.

The second boundary is thread ownership. Godot scene-tree objects, UI nodes, and
many game objects are frame-owned. The HTTP server runs on background threads.
The bridge has to cross from transport code into frame-owned state with a clear
handoff. \texttt{BridgeMod.cs} uses a \texttt{ConcurrentQueue<Action>} and drains
a bounded number of work items on \texttt{ProcessFrame}:

\begin{codeblock}[STS2 frame-owned work queue]
private static readonly ConcurrentQueue<Action> _mainThreadQueue = new();

private static void ProcessMainThreadQueue()
{
    int processed = 0;
    while (_mainThreadQueue.TryDequeue(out var action) && processed < 10)
    {
        try { action(); }
        catch (Exception ex) { GD.PrintErr($"[STS2 Bridge] {ex}"); }
        processed++;
    }
}

internal static Task<T> RunOnMainThread<T>(Func<T> func)
{
    var tcs = new TaskCompletionSource<T>();
    _mainThreadQueue.Enqueue(() =>
    {
        try { tcs.SetResult(func()); }
        catch (Exception ex) { tcs.SetException(ex); }
    });
    return tcs.Task;
}
\end{codeblock}

That queue is the core systems decision. It lets HTTP requests wait for a
frame-safe answer, limits per-frame bridge work, and keeps exceptions attached to
the request that caused them. State reads and actions then become explicit
runtime transactions:

\begin{codeblock}[STS2 request-to-runtime transaction]
var stateTask = RunOnMainThread(() => BuildGameState());
var state = stateTask.GetAwaiter().GetResult();

var resultTask = RunOnMainThread(() => ExecuteAction(action, parsed));
var result = resultTask.GetAwaiter().GetResult();
\end{codeblock}

The third boundary is semantic. The bridge should reveal decisions, choices, and
legal verbs, while the game remains the authority for simulation. The external
Python process keeps a shallow model: it asks for state, normalizes that state,
and sends typed actions. The game decides which cards exist, which targets are
legal, which modal screen is active, which reward can be claimed, and which room
transition happens next.

This is the philosophy behind the in-process bridge. Move close enough to the
runtime to read meaningful objects. Respect the engine's scheduling model. Export
the smallest stable vocabulary that supports planning. Leave combat resolution,
map progression, rewards, save state, and UI transitions to the original
software. The result is a hack in the productive systems sense: a new interface
boundary inserted at the point where the application already knows what the user
is deciding.

\subsubsection{Screens Become Decision States}

The hard part is the semantic lift. The game presents a run plus the currently
active room and whichever screen or overlay happens to be on top.
\texttt{BuildGameState}
therefore reads both backend game objects and frontend screen nodes. It checks
\texttt{RunManager}, \texttt{CombatManager}, \texttt{NOverlayStack},
\texttt{NMapScreen}, \texttt{NRewardsScreen},
\texttt{NCardGridSelectionScreen}, \texttt{NMerchantRoom}, and room classes such
as \texttt{CombatRoom}, \texttt{EventRoom}, \texttt{RestSiteRoom}, and
\texttt{TreasureRoom}. That is exactly the architecture we wanted to surface:
the original application state spans backend systems and human-facing UI layers.

The bridge turns that spread-out state into an agent grammar. Combat rooms become
\texttt{combat\_play}. Card reward overlays become \texttt{card\_reward}.
Post-combat rewards become \texttt{combat\_rewards}. Map screens become
\texttt{map\_select}. Events, shops, rest sites, relic choices, treasure rooms,
game-over screens, and fallback overlays each get a named state. The Python
state adapter then normalizes those raw bridge states into compact JSON records
with the fields an agent needs for the next move: cards, targets, energy,
choices, prices, relics, map nodes, prompts, and proceed flags.

The resulting design has a useful asymmetry. The bridge is allowed to know about
messy runtime details. The CLI surface should stay boring. An agent should see
\texttt{state}, \texttt{play-card}, \texttt{use-potion}, \texttt{end-turn},
\texttt{choose-map}, \texttt{claim-reward}, \texttt{pick-card-reward},
\texttt{event}, \texttt{rest}, \texttt{shop-buy}, \texttt{select-card},
\texttt{confirm-selection}, and \texttt{proceed}. Those verbs are less expressive
than the full internal runtime, by design. They are stable enough to plan with.

The value of that stability is especially important for data collection. In a
source-free black-box setting, collecting keyboard-and-mouse demonstrations is
labor intensive because the data unit is an entire human trajectory. The
collector must produce many examples, across resolutions, timings, rooms,
overlays, and partial failures, even when the underlying decision is the same.
With an adapter plus a CLI, the data unit changes. Once a command has been
validated, its intended effect is stable: \texttt{choose-map} chooses a map node,
\texttt{claim-reward} claims a reward, and \texttt{proceed} advances when the
game permits progress. New trajectories can then be generated by automated
exploration over command sequences and state predicates, rather than by
repeatedly recording human motor behavior. The bridge turns data collection from
imitation of pixels and input traces into composition over confirmed semantic
operators.

\subsubsection{Actions Cross the Frontend/Backend Boundary}

Action execution shows why this is a harder example than a file-backed creative
tool. Some actions can use game-domain methods directly. Others have to operate
through screen nodes and UI controls because the game has encoded part of the
operation in the interface layer. The bridge code finds menu and screen nodes,
reads private fields with reflection when needed, and calls \texttt{ForceClick}
on internal buttons and hitboxes. Starting a run, abandoning a run, choosing map
nodes, confirming card selections, claiming rewards, leaving rooms, and handling
treasure all pass through this mixed frontend/backend surface.

This is a useful lesson for agent-native interfaces. A mature software backend
makes the lift easier. Many real applications expose only partial backend
structure. Games are a clear example because the authoritative state is live and the action
system is tied to screens, animations, modal overlays, and frame scheduling. The
right harness boundary may be a small in-process bridge that reads internal
objects, respects the runtime thread model, and exports a narrow set of stable
verbs.

\begin{table}[!tbp]
\centering
\small
\begin{tabularx}{\linewidth}{>{\RaggedRight\arraybackslash}p{0.26\linewidth}YY}
\toprule
\tablehead
\textbf{Runtime problem} & \textbf{Bridge-side work} & \textbf{Agent-facing result} \\
\midrule
State spread across run objects, rooms, overlays, and UI screens &
Read game systems and screen nodes inside \texttt{BuildGameState}. &
Expose \MetricStsTwoDecisionStates{} normalized decision states through \texttt{state}. \\
Game objects owned by the frame thread &
Route HTTP work through \texttt{RunOnMainThread} and \texttt{ProcessFrame}. &
Make state reads and actions callable from a normal local CLI without racing the runtime. \\
Actions encoded in both game commands and UI nodes &
Use internal methods, screen traversal, reflection, and \texttt{ForceClick} where the game requires it. &
Expose stable verbs such as \texttt{play-card}, \texttt{choose-map}, \texttt{rest}, and \texttt{proceed}. \\
Human screens too detailed for planning &
Collapse screen-specific details into task-level choices, prompts, cards, enemies, rewards, and flags. &
Give agents a compact JSON state that supports long-horizon play and recovery. \\
Regression risk from a live, versioned game runtime &
Keep \MetricStsTwoUnitTests{} unit tests and \MetricStsTwoEndToEndTests{} E2E tests around adapter behavior and bridge workflows. &
Track the contract through tests and documented adapter behavior. \\
\bottomrule
\end{tabularx}
\caption{How the Slay the Spire II harness builds an agent-native boundary
inside a live game whose useful state spans backend objects and frontend screens.}
\label{tab:sts2-case-study}
\end{table}

The broader lesson is that \system{} has at least two growth modes. In the
Blender mode, the harness lifts a mature backend API into a stateful CLI. In the
Slay the Spire II mode, the harness establishes a backend adapter inside the
application first, then lifts that adapter into a CLI. Both routes preserve the
same agent-facing contract: explicit state, typed actions, backend truth, and
testable behavior. The difference is where the interface boundary has to be
found.

This case also reframes games as an interface benchmark. Playing a game is
normally an entertainment activity, designed around human perception, attention,
and motor control. Yet the same activity can be automated once the software
exposes explicit state and typed actions. That suggests a broader design rule for
future applications, not only for games: the GUI should remain the human-facing
surface, optimized for inspection, exploration, and enjoyment, while a CLI or
equivalent machine-facing interface should expose the same task in a compact,
testable, and compositional form for agents. The two surfaces do not compete.
They serve different users of the same software: humans need an interface they
can see and manipulate; agents need an interface they can parse, plan over, and
call reliably.

\section{Evaluation}

We evaluate \system{} as an interface system. The relevant questions are about
surface area, production usage, and ecosystem maturity: how many interfaces
exist, how broad the catalog is, how \hub{} is actually used, and whether
repository structure supports agent discovery. This emphasis follows
agent-environment evaluations
that score task completion through executable software states. Coding tasks use
tests and repositories~\cite{swebench,sweagent}; web and enterprise tasks use
browser or service state~\cite{webarena,mind2web,workarena}; app-world and
tool-user tasks check database or policy state~\cite{appworld,taubench}; and
mobile or desktop tasks inspect device and OS state~\cite{androidworld,osworld}.

\subsection{Manipulation Artifacts From Recorded Runs}

\begin{figure}[H]
\centering
\begin{minipage}[t]{0.32\linewidth}
  \begin{tcolorbox}[
    enhanced,
    colback=white,
    colframe=ReportLine,
    boxrule=0.45pt,
    arc=1.5mm,
    left=0.9mm,
    right=0.9mm,
    top=0.9mm,
    bottom=0.9mm,
    drop fuzzy shadow=ReportLine!45
  ]
    \includegraphics[width=\linewidth]{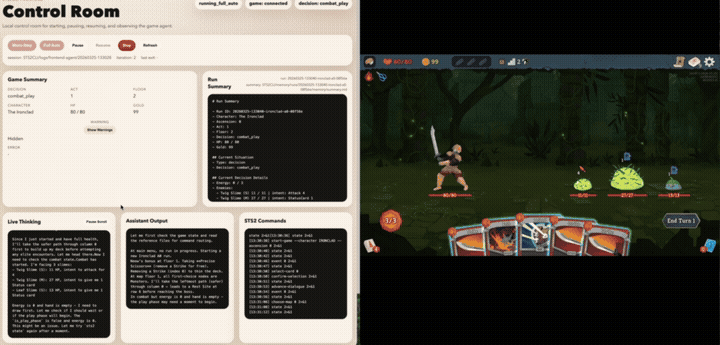}
  \end{tcolorbox}
  {\sffamily\footnotesize\bfseries Slay the Spire II\par}
  {\footnotesize\color{ReportMuted}A live game state paired with a control-room
  surface, current decision label, command log, and agent-readable run summary.}
\end{minipage}\hfill
\begin{minipage}[t]{0.32\linewidth}
  \begin{tcolorbox}[
    enhanced,
    colback=white,
    colframe=ReportLine,
    boxrule=0.45pt,
    arc=1.5mm,
    left=0.9mm,
    right=0.9mm,
    top=0.9mm,
    bottom=0.9mm,
    drop fuzzy shadow=ReportLine!45
  ]
    \includegraphics[width=\linewidth]{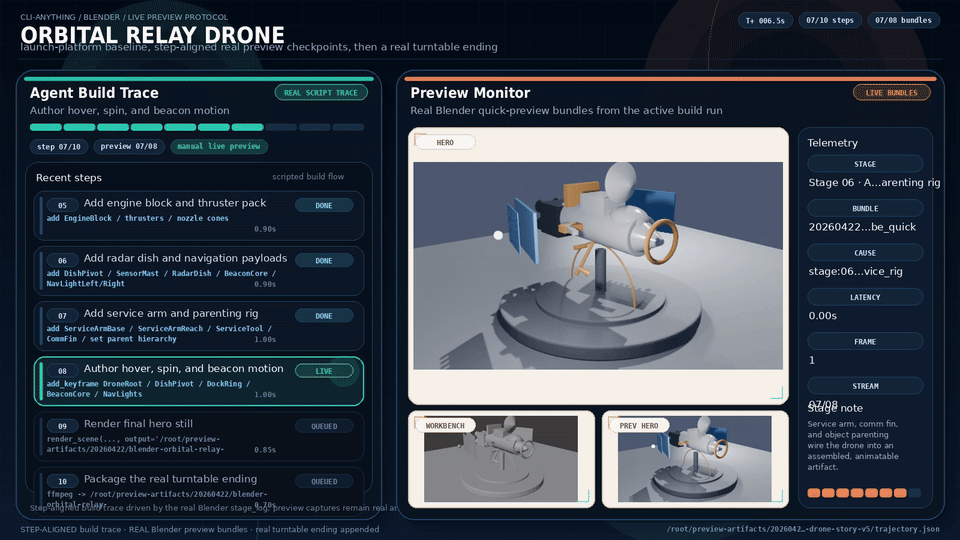}
  \end{tcolorbox}
  {\sffamily\footnotesize\bfseries Blender\par}
  {\footnotesize\color{ReportMuted}A mid-build orbital relay drone preview, with
  command progress and real Blender-rendered artifacts in the same trajectory.}
\end{minipage}\hfill
\begin{minipage}[t]{0.32\linewidth}
  \begin{tcolorbox}[
    enhanced,
    colback=white,
    colframe=ReportLine,
    boxrule=0.45pt,
    arc=1.5mm,
    left=0.9mm,
    right=0.9mm,
    top=0.9mm,
    bottom=0.9mm,
    drop fuzzy shadow=ReportLine!45
  ]
    \includegraphics[width=\linewidth]{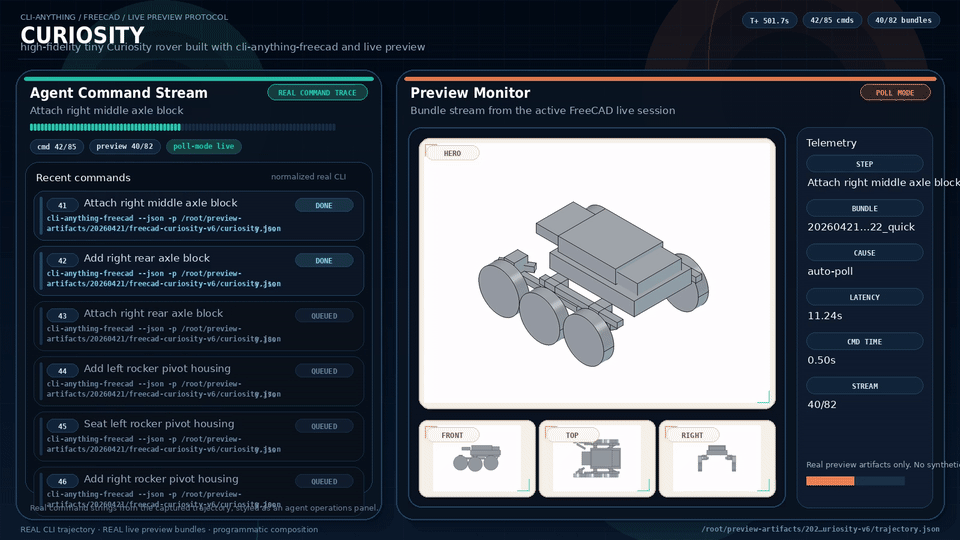}
  \end{tcolorbox}
  {\sffamily\footnotesize\bfseries FreeCAD\par}
  {\footnotesize\color{ReportMuted}An intermediate rover assembly where the
  command stream, bundle stream, and CAD preview stay aligned.}
\end{minipage}
\caption{Midpoint frames from real manipulation demos. Each frame is extracted
from a recorded run and shows a live or previewed task state tied to command
history, rather than a standalone visual mockup.}
\label{fig:evaluation-real-manipulations}
\end{figure}

Counts are only part of the evaluation story for an interface layer. A harness
also has to survive the moment where an agent changes software state and the
system produces an inspectable visual artifact. Figure~\ref{fig:evaluation-real-manipulations}
shows midpoint frames extracted from three recorded manipulation demos:
Slay the Spire II gameplay control, Blender orbital relay drone preview, and
FreeCAD Curiosity rover preview. These are not mock screenshots. They are static slices
of runs where a command surface, state/trajectory display, and rendered or live
application state move together.

The three examples stress different parts of the same contract. Slay the Spire
II validates an in-process bridge for a running game whose useful state is split
across game objects and screen state. Blender validates a backend-rendered
creative workflow, where each preview is tied to a generated scene contract and
real renderer output. FreeCAD validates a CAD trajectory where command progress,
preview bundles, and intermediate geometry remain synchronized. The evaluation
point is that agent-native interfaces need visible evidence, but that evidence
should be attributable to command/state transitions rather than detached from
the underlying backend.

\subsection{Catalog Coverage}

The current catalog has moved beyond a small demonstration set. The generated
and curated harness registry contains \MetricHarnessEntries{} entries, while the
public registry adds \MetricPublicEntries{} third-party CLIs with heterogeneous
install strategies. Together they cover \MetricCombinedEntries{} CLI entries
across \MetricCombinedCategories{} combined categories. This matters because the argument
for agent-native interfaces depends on breadth: a single harness can show that a
GUI application can be lifted; a catalog shows that the method generalizes
across creative tools, office artifacts, AI services, browser automation,
debugging, graphics, video, scientific tools, DevOps, and knowledge workflows.
The breadth goal is analogous to web and desktop-agent benchmarks that evaluate
agents across many sites, applications, and task families. Early web benchmarks
studied open-domain and shopping interactions~\cite{worldofbits,webshop};
later web benchmarks added realistic sites and enterprise
workflows~\cite{webarena,mind2web,workarena}. Visual and mobile benchmarks add tasks
where screenshots and gestures are central~\cite{visualwebarena,androidworld,
appagent}; desktop benchmarks add operating-system state~\cite{osworld}.

\begin{table}[!tbp]
\centering
\small
\begin{tabularx}{\linewidth}{YrrY}
\toprule
\tablehead
\textbf{Registry surface} & \textbf{Entries} & \textbf{Categories} & \textbf{Source} \\
\midrule
Generated/curated harness CLIs & 65 & 29 & \texttt{registry.json} \\
Public third-party CLIs & 18 & 11 & \texttt{public\_registry.json} \\
Combined catalog & 83 & 32 & both registries \\
\bottomrule
\end{tabularx}
\caption{Current CLI catalog size and category coverage.}
\label{tab:evaluation-catalog}
\end{table}

\begin{table}[!tbp]
\centering
\small
\begin{tabularx}{\linewidth}{Yrrr}
\toprule
\tablehead
\textbf{Top combined categories} & \textbf{Harness CLIs} & \textbf{Public CLIs} & \textbf{Total} \\
\midrule
ai & 6 & 3 & 9 \\
devops & 4 & 3 & 7 \\
web & 4 & 3 & 7 \\
graphics & 5 & 0 & 5 \\
video & 5 & 0 & 5 \\
communication & 2 & 2 & 4 \\
office & 4 & 0 & 4 \\
gamedev & 3 & 0 & 3 \\
image & 3 & 0 & 3 \\
knowledge & 2 & 1 & 3 \\
3d & 2 & 0 & 2 \\
audio & 1 & 1 & 2 \\
\bottomrule
\end{tabularx}
\caption{Top catalog categories by combined CLI count.}
\label{tab:evaluation-categories}
\end{table}

\begin{figure}[t]
\centering
\framedgraphic[0.92\linewidth]{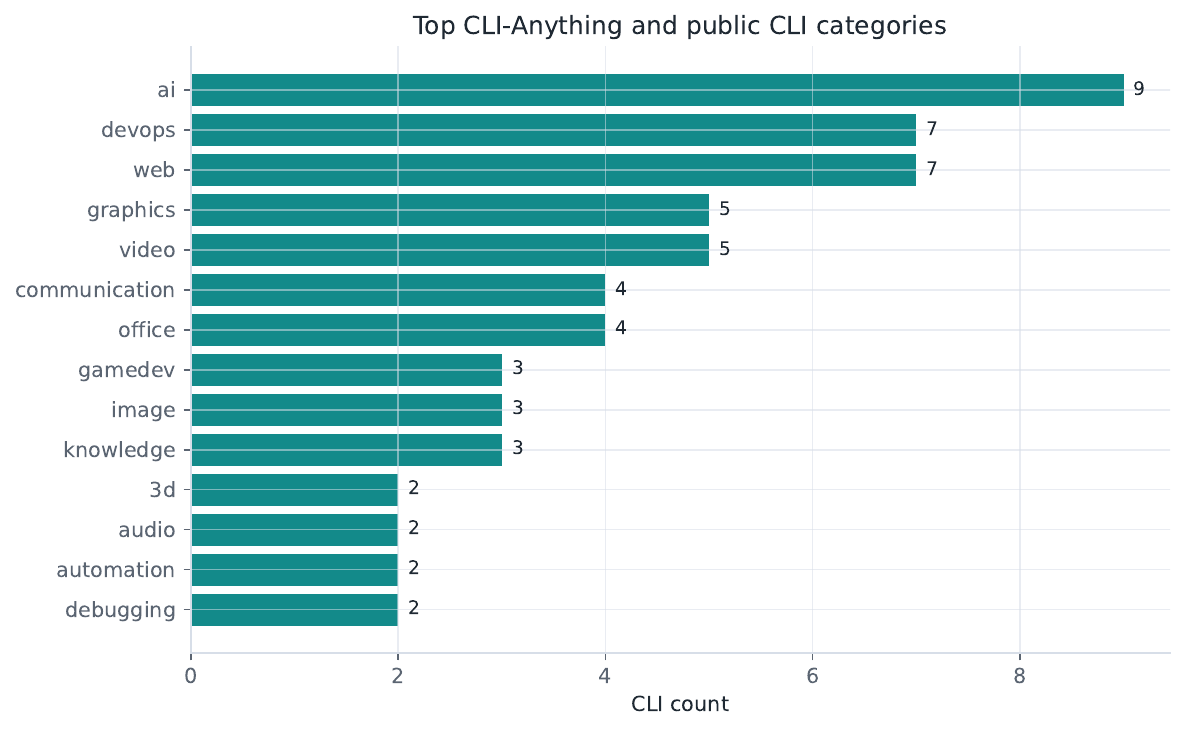}
\caption{Top combined categories across the harness and public CLI registries.
The current catalog spans AI, web, DevOps, video, graphics, office, image, and
communication tools.}
\label{fig:evaluation-categories}
\end{figure}

\subsection{\hub{} Usage}

\hub{} usage is already agent-heavy. Observed \hub{} telemetry shows an
\MetricAgentShare{} agent share and a \MetricAgentRatio{} agent/human call
ratio, and the CLI usage trend is rising rapidly. The important signal is not a
single absolute counter. It is that agents already call \hub{} as an operational
dependency, while humans still browse, install, and inspect tools through the
same distribution layer. This mixed pattern changes the design target: structured
search, install diagnostics, direct skill fetch, and preview metadata should be
treated as runtime affordances for agents, not just documentation conveniences
for people.

\begin{figure}[H]
\centering
\framedgraphic[0.95\linewidth]{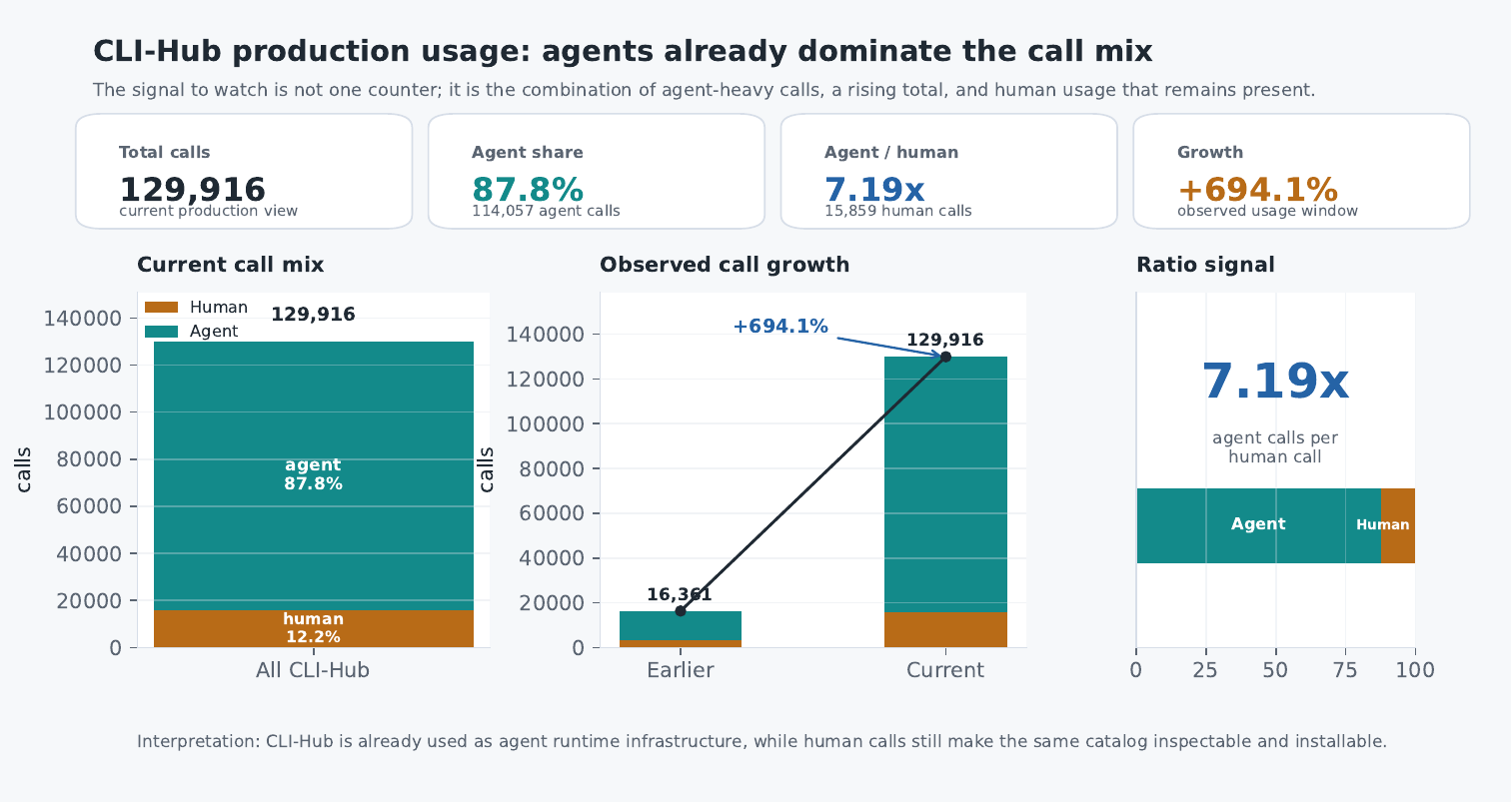}
\caption{\hub{} usage mix and trend. Agent calls dominate the current call mix,
the total call volume has grown substantially over the observed usage window,
and the agent/human call ratio shows that \hub{} is already
used as runtime infrastructure rather than only as a human package browser.}
\label{fig:evaluation-cli-hub-usage}
\end{figure}

\subsection{Additional Repository Signals}

Other support signals show the same pattern. The repository pairs CLIs with
\MetricSkillFiles{} companion skills, making the interfaces directly
discoverable by skill-aware agents. Preview-bundle integrations show that the
display protocol is being adopted beyond a single case study. The public registry
also exercises multiple install strategies: npm, pip, bundled/system CLIs, uv,
brew, and script installers. This diversity is important because a serious agent
interface layer cannot assume one package manager or one software backend.

\begin{table}[H]
\centering
\small
\begin{tabularx}{\linewidth}{Yr}
\toprule
\tablehead
\textbf{Additional agent-support signal} & \textbf{Count} \\
\midrule
Companion skills for CLIs & 61 \\
Preview bundle helper integrations & 5 \\
Public registry install strategy: npm & 8 \\
Public registry install strategy: pip & 3 \\
Public registry install strategy: brew & 2 \\
Public registry install strategy: bundled & 2 \\
Public registry install strategy: command & 1 \\
Public registry install strategy: script & 1 \\
Public registry install strategy: uv & 1 \\
\bottomrule
\end{tabularx}
\caption{Companion skills, preview integrations, and install-strategy diversity for the current CLI catalog.}
\label{tab:evaluation-other}
\end{table}

\section{Cross-Harness Lessons and Failure Modes}

The most useful lessons came from real software categories across many canonical
apps. Creative tools such as GIMP, Blender, Inkscape, Krita, and FreeCAD show
why real renderers matter. Effects, cameras, lights, fonts, image
filters, and geometry kernels often behave differently from simplified
reimplementations. Office tools such as LibreOffice and Zotero show the value of
structured package formats and headless export. Video tools such as Shotcut,
Kdenlive, Openscreen, and VideoCaptioner expose timecode, duration, codec, and
filter-mapping issues that only appear under real media validation. Debugging and
graphics tools such as LLDB, RenderDoc, Nsight Graphics, and Unreal Insights
show that inspection-heavy software can be especially agent-native once state is
surfaced as JSON and diffable artifacts.

Several failure modes recur across harnesses. First, a backend process may exit
successfully while producing no useful output. The fix is artifact validation
over exit-code trust. Second, a project-only mutation can fail to survive export
because the render path lacks a mapping for a filter or effect. The fix is to
document project-only features or add real render mappings. Third, stateless
commands lose context across turns. The fix is explicit session state,
auto-save, and status probes. Fourth, preview screenshots can become decorative
while losing truthfulness. The fix is a bundle protocol tied to real project
state and backend-generated artifacts. Fifth, installability is part of
usability. The fix is a registry, install strategies, skills, and clear backend
error messages.

These failures are mundane systems failures with higher stakes for agents than
for humans. A human can notice that an exported file is blank. An agent may
only see a path and a zero exit code unless the harness checks. A human can
remember which project window is active. An agent needs that fact in a session
file. A human can visually compare two previews. An agent needs the preview
bundle, summary, metrics, and trajectory reference.

\section{Limitations and Future Work}

\system{} has a bounded scope. Some software exposes only a compiled binary,
opaque state, or interaction patterns that are difficult to lift without
accessibility APIs or GUI automation. Some creative decisions
still require rich visual judgment. Some preview integrations are early and still
have incomplete coverage for meaningful visual state. Work on web,
smartphone, and desktop agents remains directly relevant for these cases because
some tasks still require operating through visual or browser-native interfaces.
Web-agent work covers browser-native interaction when the browser is the real
environment~\cite{worldofbits,webshop}. Later web benchmarks make those tasks
more realistic and workflow-oriented~\cite{webarena,mind2web,workarena}. Visual
and mobile-agent benchmarks cover cases where screen state and gestures remain
central~\cite{visualwebarena,androidworld,aitw}; AppAgent studies smartphone-app
operation through multimodal control~\cite{appagent}. OSWorld covers the
desktop version of the same limitation~\cite{osworld}. Older
GUI automation work remains relevant when screenshots are the only reliable
interface description~\cite{sikuli,prefab}.

The next research step is benchmarkable artifact construction. If a harness can
construct, inspect, render, and verify artifacts, then task suites can ask agents
to produce a diagram, report, CAD model, video edit, GPU-capture analysis, or
office document and then evaluate the result programmatically. This would turn
many GUI computer-use tasks back into executable environment tasks, aligning
with related work on executable coding and synthesized tool-use
environments~\cite{qwen-coder-next,deepseek-v32}. App-world, tool-user, and mobile benchmarks
provide nearby evaluation models for stateful software tasks~\cite{appworld,
taubench,androidworld}.

\section{Conclusion}

\system{} argues that the future of agents using software depends on richer
software affordances for agents, including better screen control where visual
operation remains necessary. Real applications already contain backends, project
formats, renderers, command systems, APIs, and inspectable state. The practical
work is to expose those layers as executable, inspectable, replayable contracts
while preserving the real software underneath.

Coding agents work partly because their environments execute: commands run,
tests fail, files diff, and success can be checked. Professional software needs
the same class of interface layer. If an artifact can be checked by code, it can
often be built through code. The GUI should remain available to humans and as a
fallback. For many agent tasks, the default boundary should be a stateful
contract above the application: create with commands, inspect with JSON, render
with the real backend, preview with a truthful bundle, discover through skills
and \hub{}, and verify the artifact that the user will actually open.

\end{document}